%% file: Veltair.tex
\documentclass[pageno]{jpaper}

\usepackage[normalem]{ulem}
\usepackage{graphicx}
\usepackage{amssymb}
\usepackage{algorithm}  
\usepackage{amsmath,amssymb,amsfonts}
\usepackage{algorithmicx}  
\usepackage{algpseudocode} 
\usepackage{xcolor}
\usepackage{enumitem}
\usepackage{lipsum}
\usepackage{amsmath}
\usepackage{bbm}

\usepackage{multirow}
\usepackage{multicol}
\newcommand\blfootnote[1]{%
  \begingroup
  \renewcommand\thefootnote{}\footnote{#1}%
  \addtocounter{footnote}{-1}%
  \endgroup
}

\newcommand{\Fig}[1]{Fig.~\ref{#1}}

\newcommand{\Tbl}[1]{Tbl.~\ref{#1}}
\newcommand{\Sec}[1]{Sec.~\ref{#1}}

\newcommand{\Agl}[1]{Agl.~\ref{#1}}

\newcommand{\proj}{\textsc{Veltair}}
\newcommand{\paragraphbold}[1]{\paragraph{\textbf{#1}}}

\begin{document}

\title{
VELTAIR: Towards High-Performance Multi-tenant Deep\\Learning Services via Adaptive Compilation and Scheduling}
\date{}
\author{Zihan Liu, Jingwen Leng*, Zhihui Zhang, Quan Chen, Chao Li, Minyi Guo*\\
        Emerging Parallel Computing Center, Shanghai Jiao Tong University\\
        Shanghai Qi Zhi Institution\\
        $\lbrace$altair.liu, leng-jw, zhihui-zhang$\rbrace$@sjtu.edu.cn, $\lbrace$chen-quan, lichao, guo-my$\rbrace$@cs.sjtu.edu.cn
        }




\maketitle
\newcommand{\PBox}[1]{\vspace*{.05cm}\noindent\fbox{\parbox{\columnwidth}{\vspace*{.05cm}{#1}}}\vspace*{.05cm}}
\newcommand\numberthis{\addtocounter{equation}{1}\tag{\theequation}}

\blfootnote{$*$Jingwen Leng and Minyi Guo are the corresponding authors of \proj{}}
\input{abstract.tex}

\input{Introduction.tex}
\input{Background.tex}
\input{Motivation.tex}

\input{DesignAndImplementation.tex}
\input{Evaluation.tex}

\input{RelatedWork.tex}
\input{Conclusion.tex}
\bibliographystyle{ieeetr}
\bibliography{references}

\end{document}

%% file: abstract.tex
\begin{abstract}

Deep learning (DL) models have achieved great success in many application domains.
As such, many industrial companies such as Google and Facebook have acknowledged the importance of multi-tenant DL services.
Although the multi-tenant service has been studied in conventional workloads, it is not been deeply studied on deep learning service, especially on general-purpose hardware. 
    
In this work, we systematically analyze the opportunities and challenges of providing multi-tenant deep learning services on the general-purpose CPU architecture from the aspects of scheduling granularity and code generation. 
We propose an adaptive granularity scheduling scheme to both guarantee resource usage efficiency and reduce the scheduling conflict rate. 
We also propose an adaptive compilation strategy, by which we can dynamically and intelligently pick a program with proper exclusive and shared resource usage to reduce overall interference-induced performance loss. 
Compared to the existing works, our design can serve more requests under the same QoS target in various scenarios (e.g., +71\%, +62\%, +45\% for light, medium, and heavy workloads, respectively), and reduce the averaged query latency by 50\%.

\end{abstract}

%% file: Introduction.tex
\section{Introduction}

Deep learning (DL) models have achieved great success in the various domains including vision~\cite{ResNet,MobileNetV2,EfficientNet,GoogLeNet,Tiny-YOLOV2,SSD}, natural language processing~\cite{Bert, guan2020far}, and even graph learning~\cite{zhou2021graph,cal20zhang}. 
To meet the need of rising computation power of DL models, computer architects have proposed various hardware designs including general-purpose hardware~\cite{A100} and domain-specific architectures~\cite{Eyeriss, DianNao, CambriconX, sma_dac20, TPU, zhou2021characterizing, dualsparsetc, ptolemy_micro20, tw_sc20, gan2020lowlatency} for accelerating deep learning models for their superior energy efficiency. 

Different from the computation-heavy training process, it is difficult for the inference of a single deep learning model to fully use the hardware, which typically runs with a small batch size~\cite{AI-MT}. 
As such, sharing multiple DL models on a single hardware, i.e., multi-tenant deep learning serving, has become increasingly important~\cite{Planaria, PREMA}.
Compared to the single-tenant serving, multi-tenancy brings several challenges, including resource management and allocation, shared resource competition~\cite{BubbleUp, BubbleFlux}, tasks scheduling~\cite{Heracles,Nexus}, etc. 
For conventional multi-tenant workloads, researchers have proposed various solutions based on resource partition~\cite{Parties}, hardware isolation~\cite{CuttleSys}, and so on. 
Similarly, researchers have proposed various architectural support for multi-tenant DL serving~\cite{AI-MT, PREMA, Planaria} that leverages temporal and spatial multitasking.  

However, the multi-tenant DL serving has its unique challenges, which are overlooked by previous multi-tenant DL serving works. 
We first find that owing to the complex inner-structure of the DL models~\cite{LazyBatch}, the scheduling granularity has a profound impact on the multi-model serving throughput.
Meanwhile, we demonstrate that the performance of DL models is very sensitive to code generation strategies~\cite{TVM, FlexTensor, Ansor, FAIR-TC}.
In specific, those current DL compilers mainly focus on optimizing the performance of a single model or even a single layer by various code transformations under the assumption of singe-tenancy.
Our experimental results show that the performance of generated code degrades rapidly under multi-tenant scenarios due to the shared resource contention.


In this work, we propose \proj{}, a software solution that provides the high-throughput and low-interference multi-tenant deep learning serving. 
We systematically analyze the resource allocation conflict and inter-layer interference on the CPU platform, which closely represents the industrial practice~\cite{FBCPUDataCenter}.
Our analysis indicates that the fixed scheduling granularity adopted by previous works~\cite{Planaria,PREMA} is sub-optimal when the system load changes.
Meanwhile, we perform a naive extension to the TVM's auto-scheduler~\cite{TVM}, which lets us identify the best-performing code version under different interference levels. 
We show that the performance of the best code version under a specific interference level degrades quickly under a different interference level.
These insights call for both adaptive scheduling and adaptive compilation for achieving the high-performance multi-tenant DL serving.


For the adaptive scheduling, we find that the sub-optimal performance of the fixed model-wise scheduling scheme is caused by the inefficient CPU resource utilization, while the fixed layer-wise scheduling scheme is caused by the frequent resource conflict.
To reduce the resource conflict with CPU resource usage efficiency guaranteed under different situations, we propose a layer-block granularity scheduling strategy, which is finer than the model-wise scheduling but coarser than layer-wise scheduling. By setting a dynamic threshold, we can achieve both low conflict possibility and high CPU resource usage efficiency.

For the adaptive compilation, we analyze the relationship between the interference-prone code version and the interference-tolerant code version for a set of deep learning layers.
We find that those different versions essentially lie in the Pareto frontier of trade-off space between parallelism and locality. Given this insight, we propose a single pass compiling strategy based on the existing auto-scheduler.
The extended auto-scheduler is able to compile multiple versions of implementations that are suitable for different system interference pressure levels. 

To evaluate our design, we choose various workloads from the industry-level MLPerf~\cite{MLPerf} benchmark ranging from light to heavy workload and compare with the existing multi-tenant DL serving solution, Planaria~\cite{Planaria}.
Compared to the existing work, our design serves more requests under the same QoS target in various scenarios (e.g., +71\%, +62\%, +45\% for light, medium, and heavy workloads, respectively), and reduce the averaged query latency by 50\%. 

To summarize, we make the following contributions in this work.
\begin{itemize}
\item We analyze and identify the performance-critical optimization knobs for multi-tenant DL services, including the adaptive scheduling and the adaptive compilation (\Sec{sec:OptimizationSpace}).
\item We propose a static multi-version compiler that extends the existing TVM's compilation framework and can identify different optimal code versions under different interference levels. The key novelty in our compiler is a multi-version search algorithm in a single pass (\Sec{subsec:adaptive_compilation}).
\item We propose a runtime scheduler design that dynamically forms a layer-block as the scheduling unit. 
The scheduler uses a dynamic threshold-based layer-block formation algorithm to balance the resource usage efficiency and scheduling conflict rate (\Sec{subsec:adaptive_scheduling}).
\item We evaluate and compare the proposed ideas in \proj{}, where the combined adaptive compilation and scheduler can improve the system by 45\% - 71\% in different workload mixes. 
We also show that the query execution latency in our design is within 10\% gap of the isolated execution case, meaning \proj{} is close to the performance upper bound on the studied hardware platform (\Sec{sec:Evaluation}).


\end{itemize}


%% file: Background.tex
\begin{figure}[t]
    \centering
    \includegraphics[width=0.5\linewidth]{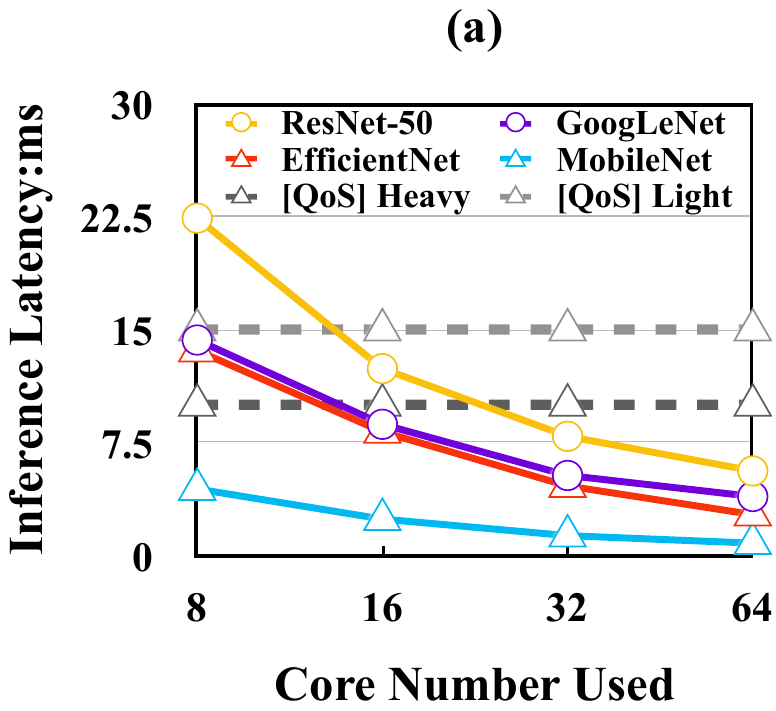}
    \includegraphics[width=0.48\linewidth]{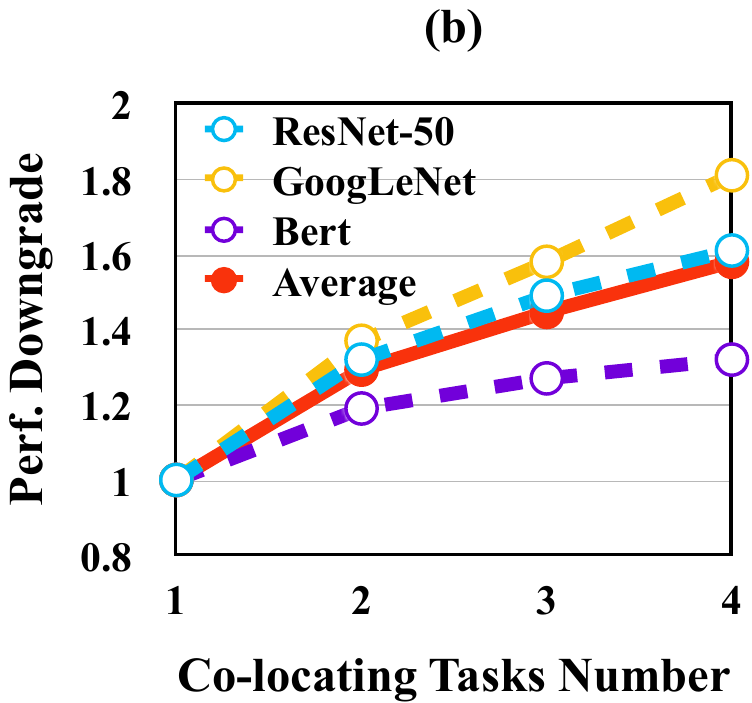}
    \caption{\label{fig-InterferenceSnapshot} (a) All models in MLPerf vision can meet the QoS target by using a few cores. (b) Performance slowdown when simply co-locating multiple tasks together. }
\end{figure}

\section{Motivation and Challenges for Multi-Tenant DNNs}
\label{sec:Background}

In this section, we explain why the CPU architecture is suitable for providing the multi-tenant DL services.
In specific, we show that the existing high-performance CPU is more than enough to serve multiple deep learning inference tasks under their QoS target.
We then show that the quality of code generation is the key to fully unleashing the potential of the underlying hardware.

\subsection{DNN Execution Characterization on CPU}
\label{subsec:execution_characterization}

With the tremendous improvement of hardware architecture design and manufacturing process, the performance of single computing hardware is increasing rapidly, making training and inference of deep neural networks easier and faster.
Some companies even use CPUs as their deep learning back-end. 
These hardware mostly use multi-degree parallelism to increase overall throughput. 
However, when providing deep learning inference services with small batch sizes, these hardware will suffer from severe under-utilization since the deep learning inference service is not intense enough to fill the hardware resource. 
 
As illustrated in \Fig{fig-InterferenceSnapshot}a, a high-performance CPU (AMD Threadripper 3990X~\cite{3990X}) is more than enough to provide deep learning inference tasks. When serving vision tasks in MLPerf~\cite{MLPerf}, the CPU platform can reach around 300 Query per Second by simply using all CPU cores for a task. So, to fully utilize the hardware and increase the energy efficiency, increasing deep learning service providers begin to introduce the task-level parallelism by sharing one computing hardware among multiple customers/requests which is called \textbf{multi-tenant deep learning service}, by either temporal multiplexing (e.g. PREMA~\cite{PREMA}, AI-MT~\cite{AI-MT}) or spatial multiplexing (e.g. NVIDIA Multi-Process Service~\cite{VoltaMPS}, NVIDIA Ampere Multi-Instance GPU~\cite{MIG}). 
By leveraging task-level parallelism, multiple customers/requests can fully occupy the throughput of the hardware, thus increasing the overall efficiency. 

On the other hand, some deep learning tasks also consist of multiple sub-tasks.
For example, auto-piloting on a smart vehicle consists of multiple direction object sensing and tracking tasks, SLAM tasks, decision-making tasks, etc.; personal voice assistant service on a home device consists of voice recognition, voice synthesis, etc.
Those sub-tasks can but also should be launched in a parallel way for real-time interaction, and thus may share the resources on a single computing hardware. 
Currently, few deep learning systems are designed to face the multi-tenant serving scenarios. 
So in this work, we propose to explore the optimization opportunities at both compiling and runtime for co-locating and scheduling multiple tasks on a single hardware. 
Specifically, we focus on the problem of co-locating multiple latency-critical DL tasks on a multi-core architecture hardware, and the objective is to serve as many DL tasks as possible (i.e., maximize the metric of query per second) under the task latency constraints (i.e., ensuring that it finishes within a time limit). However, our design can be easily extended to support the co-location of DL tasks and best-effort tasks.

To co-locate multiple deep learning tasks on a single computing back-end, one naive approach is to 
simply dump all the candidate tasks onto the hardware and fill the empty slot once a task is complete. 
However, the most important challenge is how to manage the limited hardware resources like physical cores for the CPU architecture,  streaming multiprocessors (SMs) for the GPU architecture, or even sub-arrays of a systolic architecture. 

In addition to exclusive resources, various shared resources on the computing back-end are critical to the performance of a task, including cache bandwidth, cache capacity, memory bandwidth, etc.
Naively scheduling all candidate tasks to the hardware would result in severe interference and performance loss due to the competition of these resources. 
We conduct a simple experiment that co-locates multiple ResNet-50, GoogLeNet, and SSD inference tasks on a single CPU. As illustrated in \Fig{fig-InterferenceSnapshot}b, the task suffers from up to $ 1.8\times $ latency under heavy workload pressure.
As such, the inference can severely impact the QoS, but is considered by current DL serving systems.
In contrast, our work considers both compilation and scheduling strategies to handle the interference. 

\begin{figure}[t]
    \centering
    \includegraphics[width=0.99\linewidth]{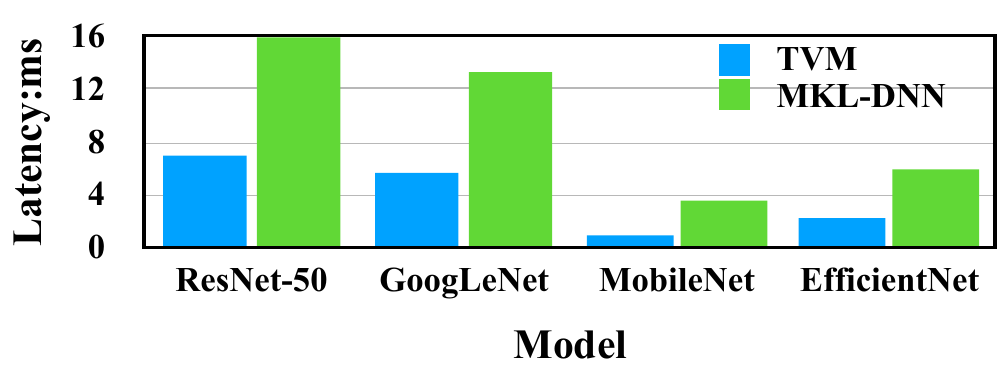}
    \caption{\label{fig-VSMKLDNN} Performance comparison between vendor-supplied MKL-DNN library and TVM compiler.}
\end{figure}

\subsection{DNN Compilation on CPU}
\label{subsec:compilation}

Although vendor-provided libraries can offer optimized DNN computation with convenient APIs, an increasing number of researchers and developers have begun to use automatic high-performance code generators for even higher performance.  
Among deep learning compilers, TVM gains great success for its convenience, high quality of generated code, and cross-platform capability. Moreover, things are getting more convenient once the TVM Auto-Scheduler (i.e., Ansor~\cite{Ansor}) is introduced. Now, researchers can simply define the computation logic they want, run the auto-scheduling procedure, and TVM would return the code with similar or even better performance compared to vendor-provided libraries, such as MKL-DNN~\cite{MKLDNN} and MLAS~\cite{MLAS} on CPU, cuDNN~\cite{cuDNN} on GPU. 

The other advantage of using the DNN compiler is that the generated code is user-visible while vendor-supplied libraries are usually closed-sourced.
Given those reasons, we choose the TVM compiler to generate the codes for running DNN models in this work.
We also conduct a performance comparison experiment between the Intel MKL-DNN~\cite{MKLDNN} and TVM.
As \Fig{fig-VSMKLDNN} shows, the TVM generally outperforms the vendor-supplied library.

For compiling DNN layers or models on the CPU, we mainly consider the nested loop transformation and some CPU-specific annotation or pragma including parallelization and unrolling. The compiling procedure is actually a trade-off between the parallelism and locality of the program, which we will discuss later in the paper.





%% file: Motivation.tex
{
\begin{table}[t]
    \setlength{\arrayrulewidth}{1.1pt}
    \renewcommand{\arraystretch}{1.4}
    \caption{Optimization strategies in \proj{} and prior works.}
    \label{tbl:work_comparison}
    \centering  
    \resizebox{\linewidth}{!}{
    \begin{tabular}{|c|c|c|c|}
        \hline
        \textbf{Multiplexing} & \textbf{Granularity} & \textbf{Compilation} & \textbf{Work}\\
        \hline
        \multirow{2}{*}{Temporal} & Static (Model) & \multirow{2}{*}{Static} & PREMA~\cite{PREMA} \\
        & Static (Layer) &  & AI-MT~\cite{AI-MT} \\\hline
        
        \multirow{4}{*}{Spatial} & Static (Model) & \multirow{2}{*}{Static} & Planaria~\cite{Planaria} \\
         & Static (Model/Layer) &  & Parties~\cite{Parties} \\\cline{2-4}
         & Static (Model/Layer) & Adaptive & Protean~\cite{Protean} \\
         & Adaptive (Layer Block) & Adaptive & \proj{} (ours)\\
        \hline
    \end{tabular}
    }
\end{table}
}

\section{Optimization Space Analysis}
\label{sec:OptimizationSpace}

In this section, we first identify the optimization space that is critical for achieving high-performance multi-tenant DL services.
In specific, we study the two optimization knobs, namely the scheduling granularity and compilation strategy.

We characterize the impact of those two knobs on the performance measured by QoS satisfaction rate~\cite{QSR, Prophet, Parties, Nexus} representing how many requests are finished within the QoS target of multi-tenant deep learning services on the CPU.  

Our main finding is two-fold. 
First, a fixed scheduling granularity, such as the entire model~\cite{PREMA} or the sub-layer block~\cite{AI-MT, Planaria}, leads to the sub-optimal performance, owing to the diversity of DNN models and their distinctive inner characteristics.
Second, the performance of the existing compilation strategies, which aim to maximize the code performance under the solo-run case, degrades significantly when multiple DNN models run together and interfere with each other. 
Such two findings motivate the design of the \emph{adaptive scheduling} and \emph{adaptive compilation} in \proj{}.

\subsection{Optimization Space Definition}

We first explain and define the optimization space of multi-tenant DL services.
Conventional workloads such as Silo~\cite{Silo} and Moses~\cite{TailBench} choose the entire query as the scheduling unit because the query has no internal structures.
In contrast, DNNs are layer-based, for which the scheduling unit can range from one layer to the entire model.
Meanwhile, DNNs are also computation-intensive, and their performances are sensitive to the code quality as shown in \Sec{subsec:compilation}.
In this work, we consider these two knobs jointly, i.e., scheduling granularity and compilation strategy, for achieving high-performance multi-tenant deep learning services.

\textbf{Scheduling granularity} refers to the size of the entity for allocating resources and scheduling on the hardware.
For example, in the conventional online services, prior works typically choose the entire query as the scheduling unit~\cite{Parties,Protean,Heracles,CuttleSys}.
However, we have more choices on the scheduling granularity in deep learning services because DNN models have a complex inner organization consisting of \emph{layers} or \emph{operators}, such as \texttt{conv} (i.e., convolution), \texttt{relu} and \texttt{pooling}. 
As such, we can either choose an entire model (i.e., coarse-grained) or a single layer (i.e., fine-grained) as the scheduling unit.  
To achieve higher resource usage efficiency and reduce the resource usage conflict, we consider a new scheduling granularity of multiple layers as a unit, which we call \emph{layer block}.

\begin{figure}[t]
    \centering
    \includegraphics[width=0.52\linewidth]{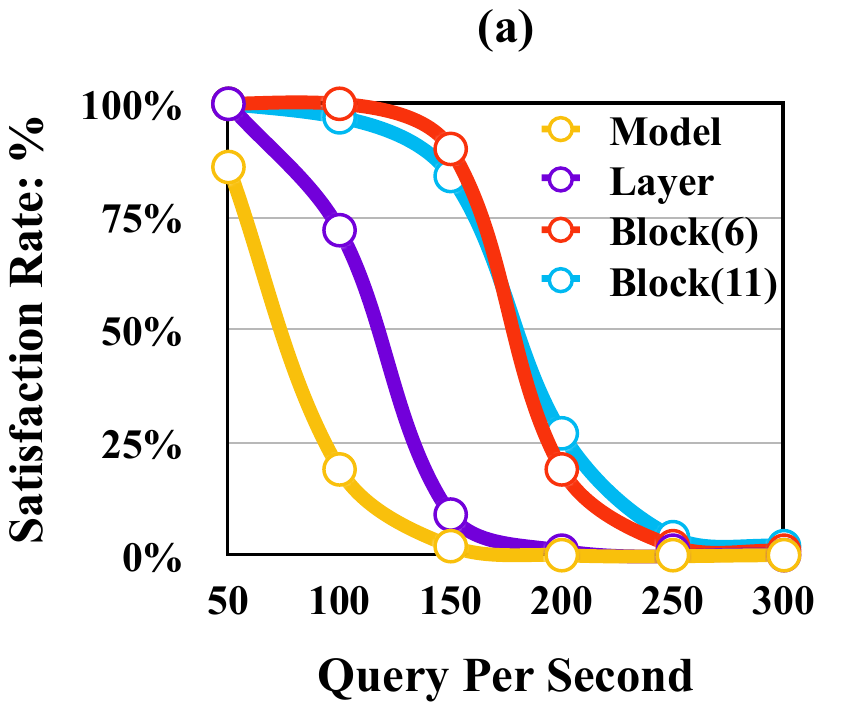}
    \includegraphics[width=0.46\linewidth]{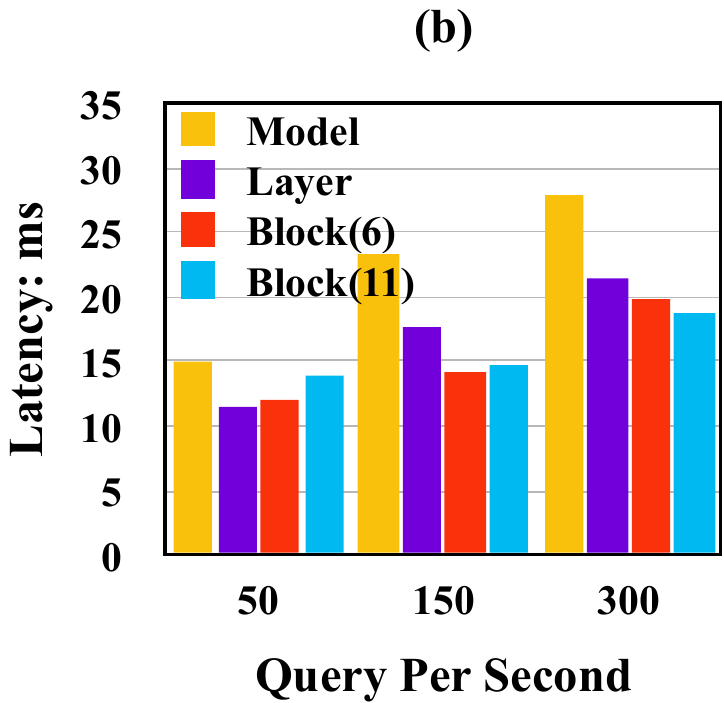}
    \caption{\label{fig-GrainPerf} Performance comparison of different scheduling granularities under different query arrival rate. (a) QoS satisfaction rate. (b) Average query latency.
    }
\end{figure}

\textbf{Compilation strategy} refers to the code generation options for a DNN model or a DNN layer.
For example, we have described the different code generation options (i.e., nested loop transformation) in \Sec{subsec:compilation}.
As we will show later, the optimal code generation option for minimizing the execution latency depends on the interference levels caused by co-running DNN models.
To the best of our knowledge, we are the first to consider the compilation as an optimization knob in the multi-tenant DNN serving scenario.

\Tbl{tbl:work_comparison} compares the choices of those two optimization knobs in prior state-of-the-art solutions against \proj{}. 
Specifically, our work is \emph{adaptive} in both the scheduling granularity and compilation strategy.
Previous work AI-MT~\cite{AI-MT} and Planaria~\cite{Planaria} decompose a layer into multiple smaller parts, or sub-layers, for more flexible scheduling.
However, the improvement is limited as we will show that the layer-wise scheduling unit is already inferior to our adaptive block scheduling in \Sec{subsec:scheduling_analysis}.
In other words, sub-layer scheduling is overly fine-grained.
Other work Protean~\cite{Protean} and Parties~\cite{Parties} mainly target conventional interactive services, so both of them use static scheduling.
The static scheduling can either choose the layer or the entire model in the case of multi-tenant DL services.
Note that Protean~\cite{Protean} also uses an adaptive compilation strategy, only targeting non-DL workloads.
As such, it can not be applied to compile DNN models because they have a different set of compiler optimization options as we will show later.
In the following parts of this section, we will justify the choice of our optimization knobs through detailed experimental results.

\subsection{Scheduling Granularity Analysis}
\label{subsec:scheduling_analysis}

We first compare the performance of multi-tenant DL services under different scheduling granularities, including layer-wise, model-wise, and layer-block scheduling. 
We then explain why these static schedulings fail to fully utilize the hardware resources, which leads to the need for an adaptive scheduling granularity.


\paragraphbold{Experimental Setup.}
For the model-wise scheduling, we implement a simple First Come First Serve (FCFS) strategy used in prior work~\cite{Clockwork, Nexus}.
In other words, the tasks will be served immediately if there are available resources while waiting otherwise. 
For the layer-wise scheduling, we implement an algorithm similar to Planaria~\cite{Planaria} that allocates the resource to every layer and allow tile-wise preemption if the requested resources exceed the available number. 
For the layer block scheduling, we simply set the layer block-size to 6 and 11 respectively to study the impact of the block-size.
We compare the performance of those scheduling schemes under different query arrival rates (i.e., query per second, QPS).
We report the QoS satisfaction ratio in \Fig{fig-GrainPerf}a and averaged model execution latency in \Fig{fig-GrainPerf}b as evaluation metrics.
For the fair comparison of different scheduling strategies, we run a total number of 30, 000 ResNet-50 models with identical uniform arriving times to eliminate the instability caused by the randomness.

\begin{figure}[t]
    \centering
    \includegraphics[width=0.495\linewidth]{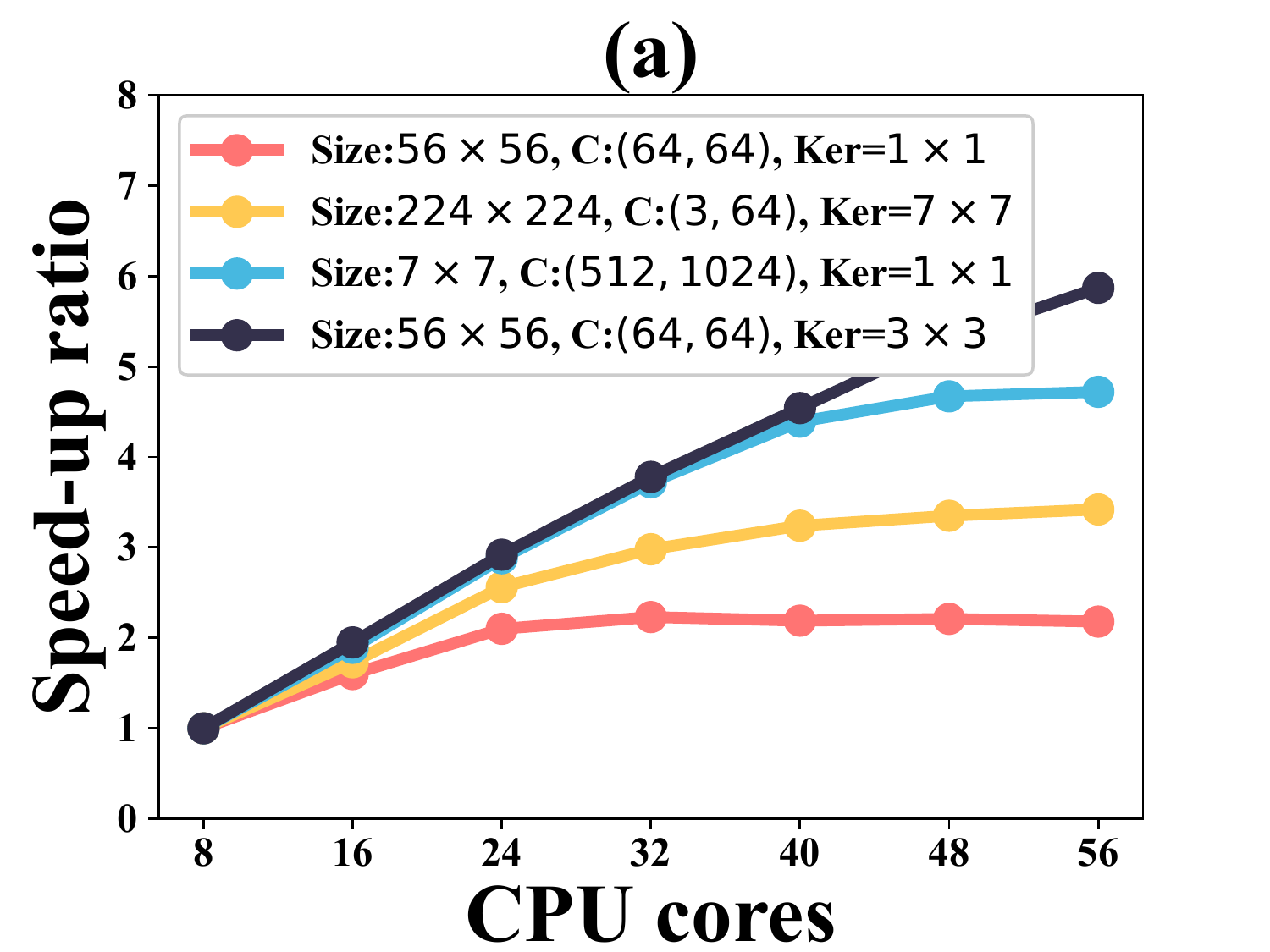}
    \includegraphics[width=0.495\linewidth]{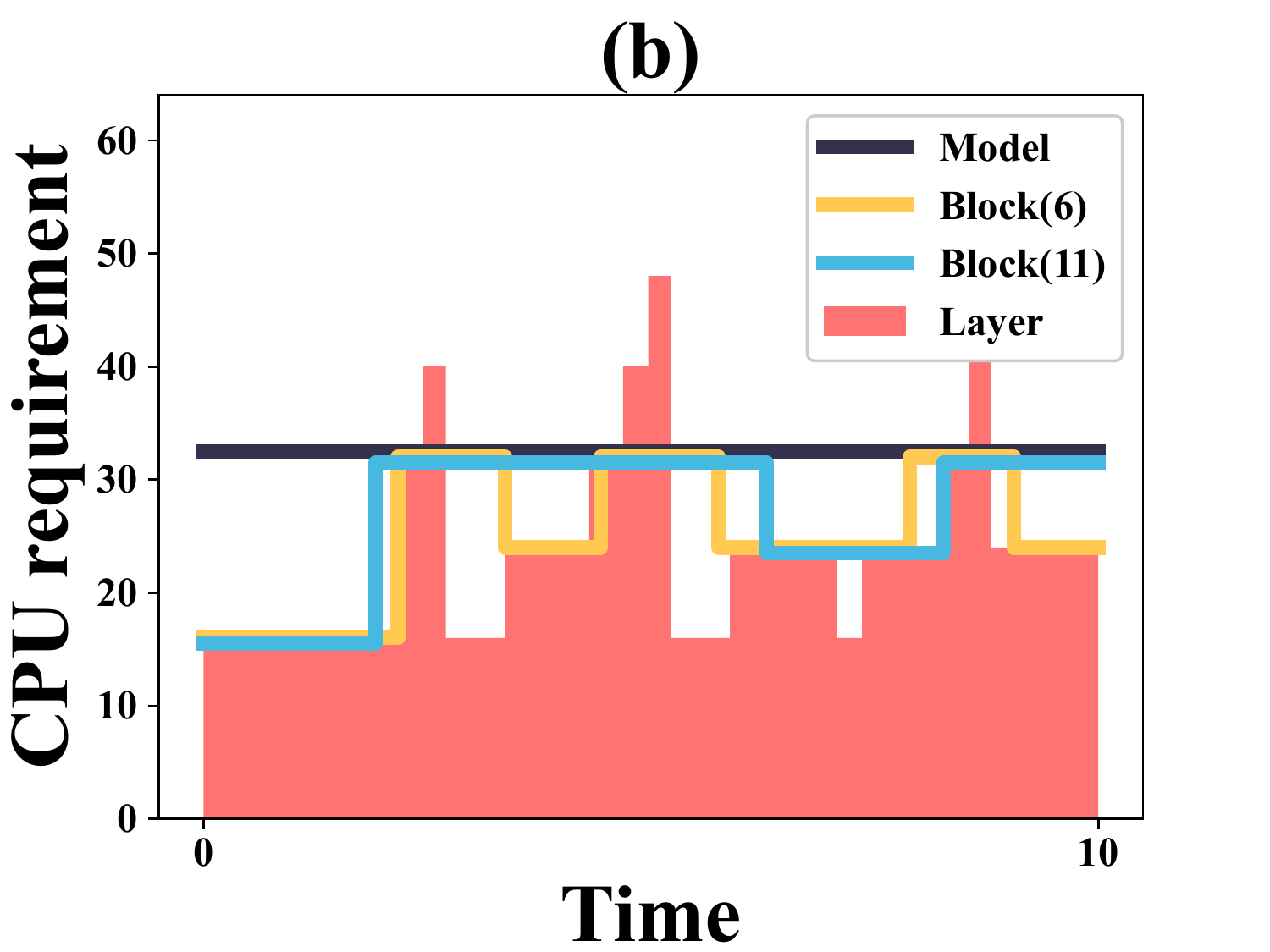}
    \caption{\label{fig-CoreUtil} (a) Speedup trend of increasing core number for 4 different \texttt{conv} layers selected from the ResNet-50 model. 
    (b) Core number allocation comparison for different scheduling granularities under the same QoS target. The layer-wise scheduling approach (red shadowed area) represents the minimum required core number for meeting the QoS target. }
\end{figure}

\paragraphbold{Results.}
As shown in \Fig{fig-GrainPerf}a, the performance of both model-wise and layer-wise scheduling degrade much faster than the block-wise scheduling.
Meanwhile, the best block-size for optimal performance varies with the query arrival rate.
For example, the block-size of 6 layers performs best at 150 QPS, while the block-size of 11 is better at 200 QPS.
We have the same observations in \Fig{fig-GrainPerf}b for the averaged query execution latency.
These results confirm the criticality of scheduling granularity for multi-tenant DL services.

\paragraphbold{Model-Wise Inefficiency.}
We find that the distinctive computation resource requirement across DNN layers is the root cause for why the model-wise scheduling is sub-optimal. 
\Fig{fig-CoreUtil}a plots the speedup for different ResNet layers under different CPU core numbers, which shows that different layers have different scalability trends when allocated core number increases.
However, the model-wise scheduling evenly assigns a fixed number of cores to all layers in the model, which results in the CPU core resource wastage because many layers only require a small number of cores.
\Fig{fig-CoreUtil}b compares the core number allocation between the model-wise scheduling and layer-wise scheduling.
Intuitively, the layer-wise scheduling scheme represents the minimum core allocation for satisfying the model's QoS target.
We find that the model-wise scheme allocation (black line) is far from the optimal core allocation (red shadowed area).
As a result, the QoS satisfaction ratio drops dramatically once the query arrival rate exceeds 50 QPS in \Fig{fig-GrainPerf}a.

\begin{figure}
    \centering
    \includegraphics[width=0.495\linewidth]{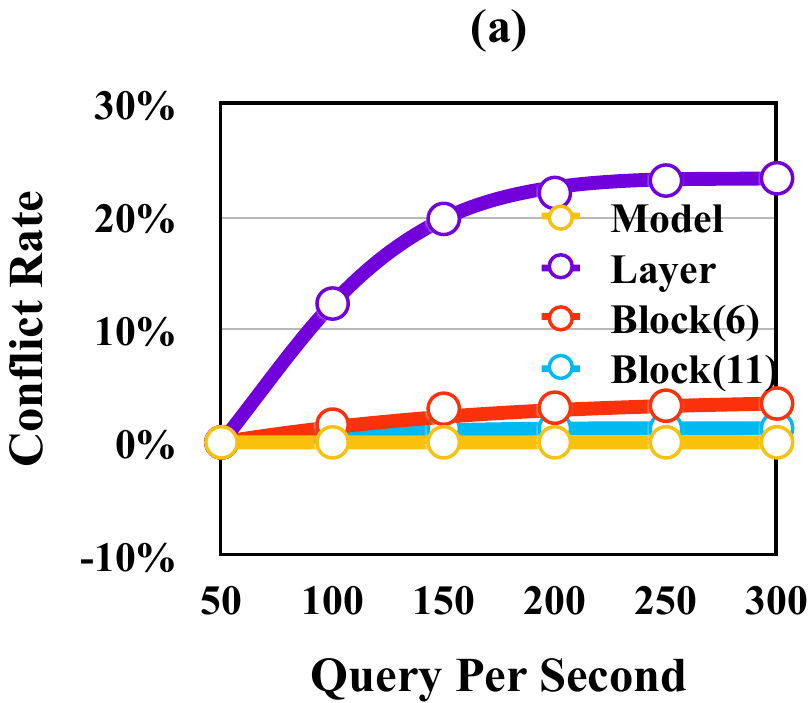}
    \includegraphics[width=0.495\linewidth]{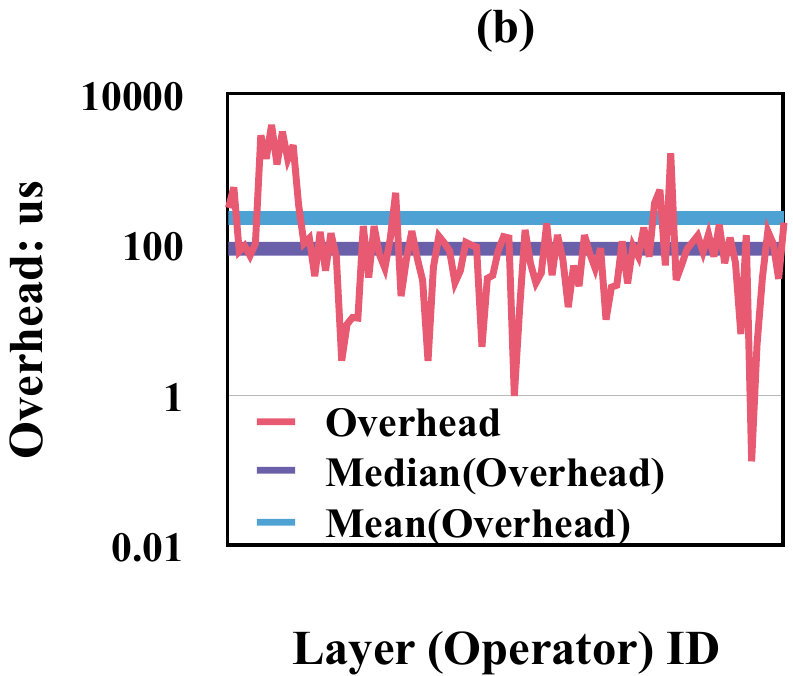}
    \caption{\label{fig-OverheadAna} (a) Scheduling conflict rate comparison for different scheduling granularities under different query arrival rates. (b) Measured per-layer conflict scheduling overhead. }
\end{figure}

\paragraphbold{Layer-Wise Inefficiency.}
We find that the layer-wise scheduling is sub-optimal owing to the frequent \emph{scheduling conflict} when the query arrival rate is high.
For example, there are layers in \Fig{fig-CoreUtil}b that require large core numbers (e.g., more than 48 out of 64 cores).
The scheduling conflict occurs when a layer requests more cores than currently available cores.
\Fig{fig-OverheadAna}a compares the conflict rate among different scheduling granularities, where the layer-wise scheduling is highest (e.g., $ 23.8\% $ conflict rate with 300 QPS).

For a layer that experiences scheduling conflict, we implement a technique to increase the resource utilization. 
In specific, we first let the layer use all the available cores and increase its core usage once more cores become available.
However, using more cores needs to spawn more threads, whose overhead is non-negligible and worsens the model's overall latency.
To illustrate this point, we quantify this overhead for each layer in ResNet-50 by measuring a layer's latency with and without scheduling conflict.
\Fig{fig-OverheadAna}b shows the results, with the mean of 220~$\mu s$ and median of 100~$\mu s$.

The scheduling conflict overhead measured above explains the overall latency of the layer-wise scheduling for ResNet-50 at the 300 QPS.
The execution latency without scheduling conflict is $18.54~ms$.
But with the conflict rate of $23.8\%$, the total conflict overhead is estimated to be $23.8\%\times 55 \times 220~\mu s = 2.86~ms$, with the 55 layers (53 \texttt{conv} and 2 \texttt{GEMM}) in ResNet-50.
As such, the estimated overall latency is $2.86 + 18.54 = 21.4 ms$, which matches the measured latency for the layer-wise scheduling at 300~QPS in \Fig{fig-GrainPerf}b.

 


\begin{figure}[t]
    \centering
    \includegraphics[width=0.495\linewidth]{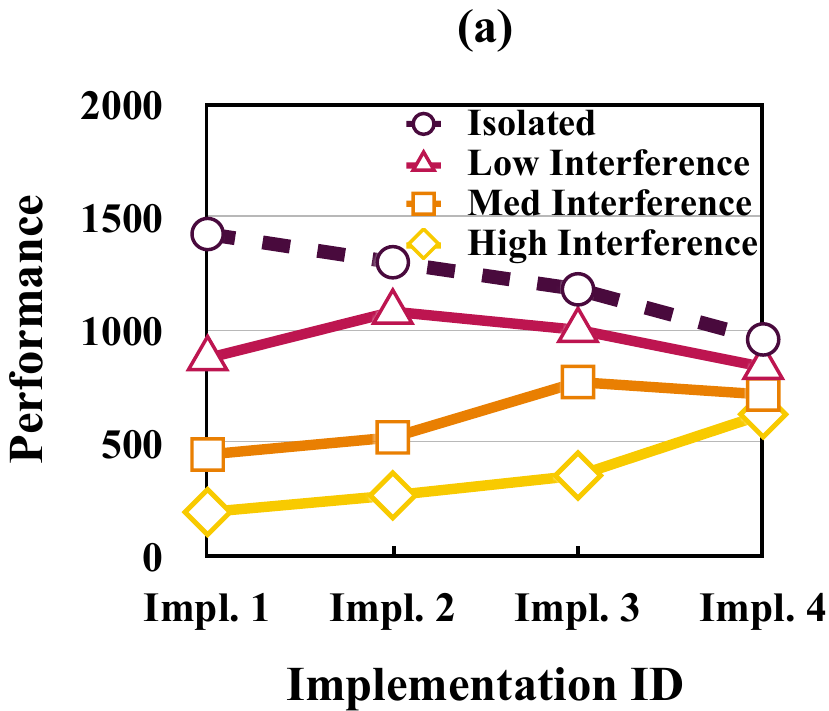}
    \includegraphics[width=0.495\linewidth]{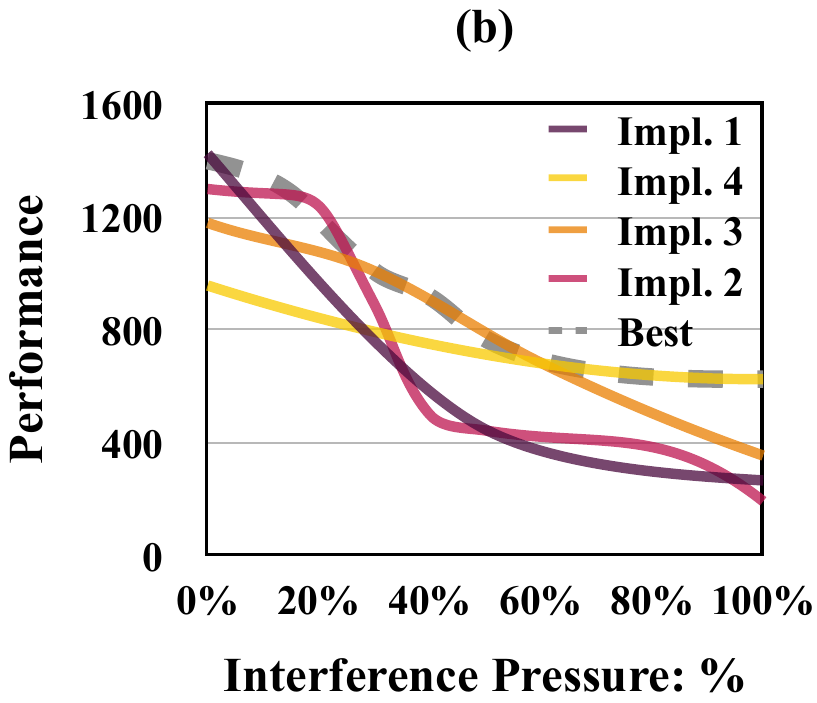}
    \caption{
    (a) Performance of four versions of the same layer under different interference levels.
    (b) The best performance (dotted line) is achieved by combining all four versions. 
    }
    \label{fig-VerBehaviourAndLossRegion}
\end{figure}

\begin{figure}[b]
    \centering
    \includegraphics[width=0.495\linewidth]{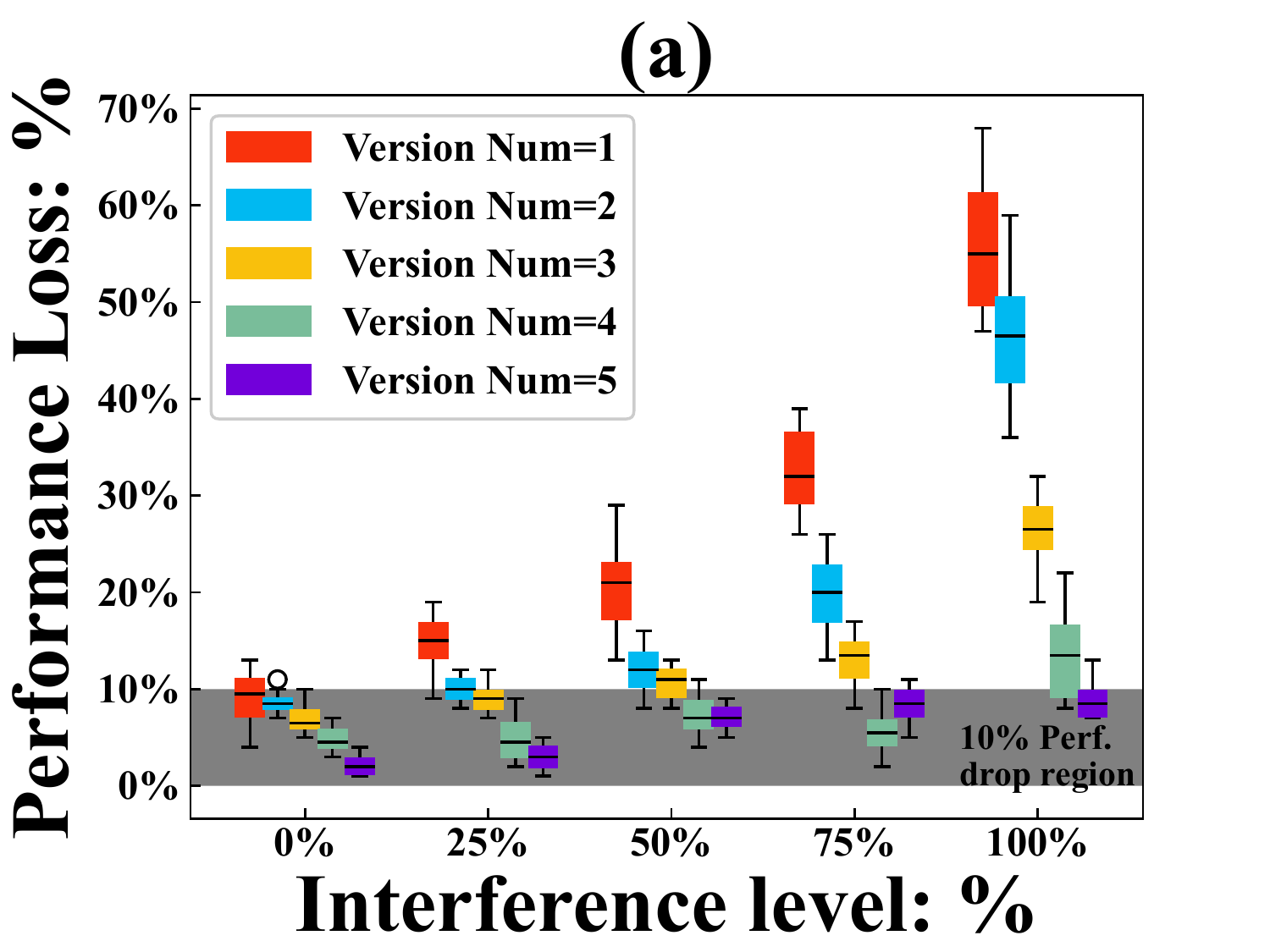}
    \includegraphics[width=0.495\linewidth]{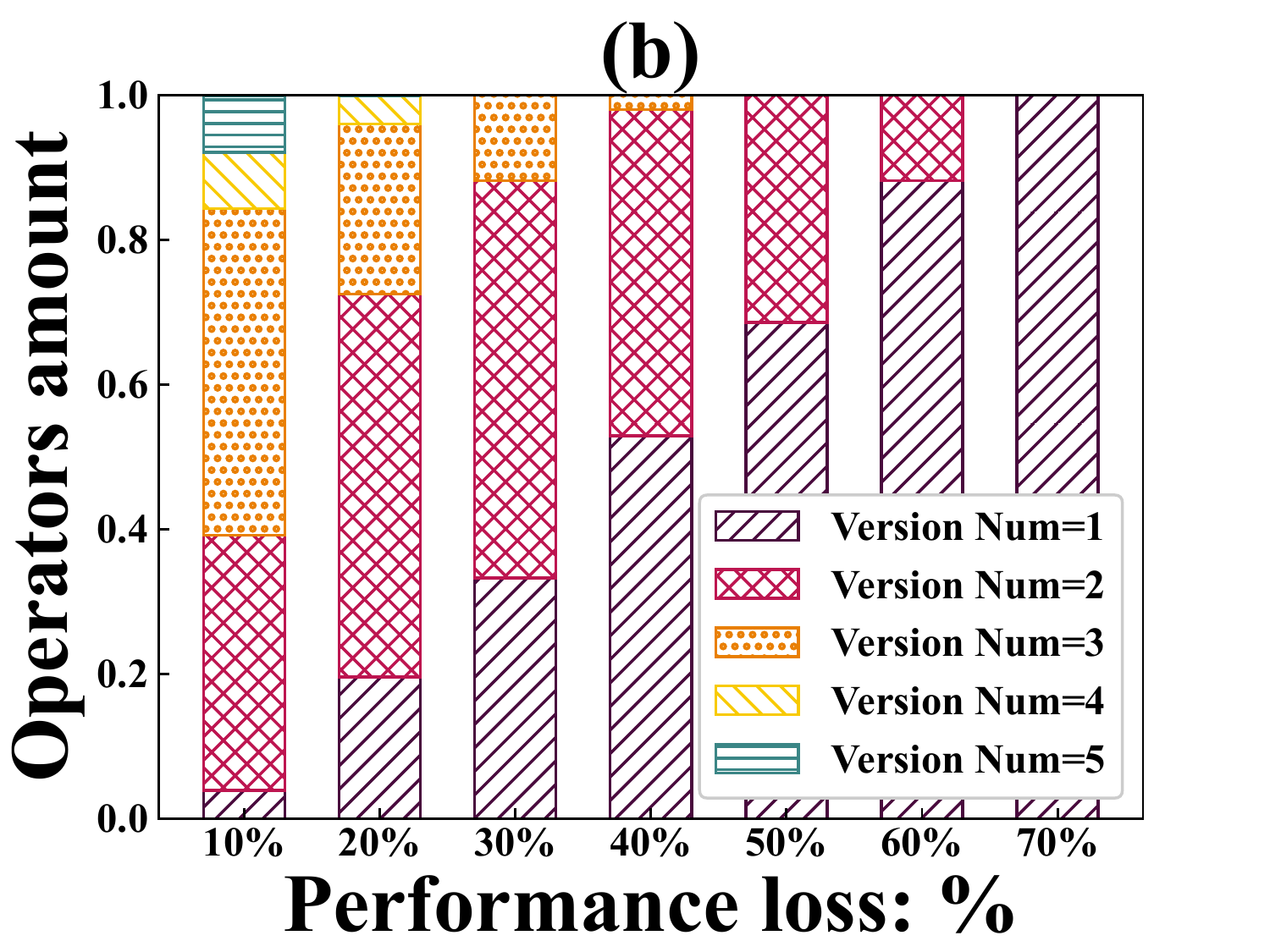}
    \caption{
        (a) Performance loss of retaining different numbers of versions compared against retaining all ten versions under different interference levels. 
        (b) Distribution of code version count to maintain various performance loss.}
        \label{fig-VerLossAndVerNum} 
\end{figure}

\begin{figure*}
    \centering
    \includegraphics[width=0.99\linewidth]{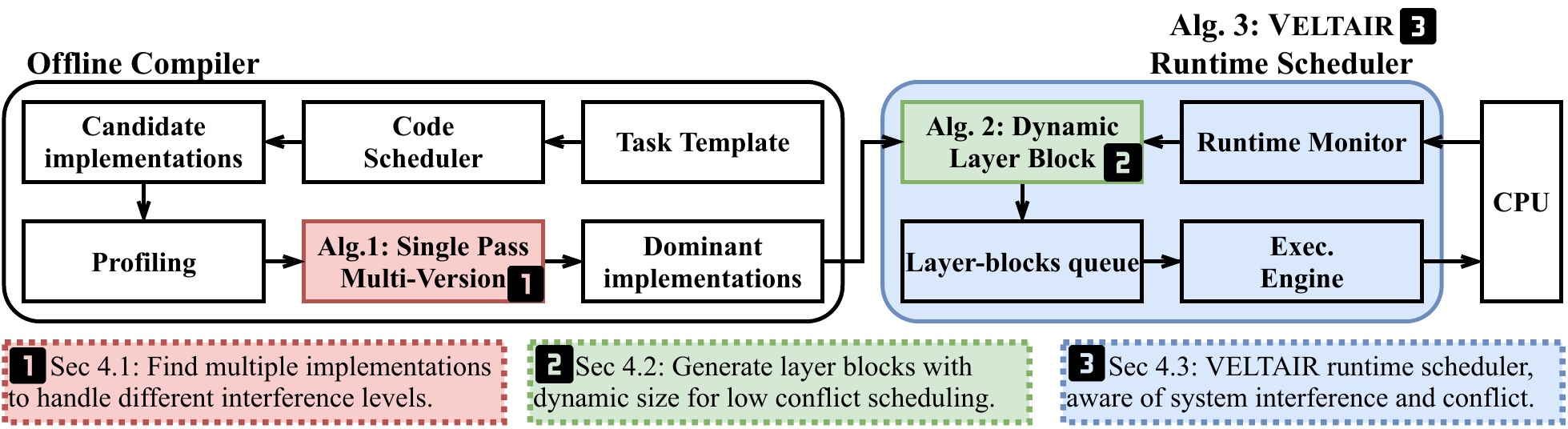}
    \caption{
    Overview of \proj{}, which comprises of the offline static compiler and the online runtime scheduler for adaptive compiling and scheduling.
The static compiler and runtime scheduler leverage the single-pass multi-version search and dynamic threshold-based layer-block formation algorithm, respectively.
By monitoring the system load and interference pressures, the scheduler adaptively selects the optimal code version and scheduling granularities.
    }
    \label{fig-workflow} 
\end{figure*}

\paragraphbold{Summary.}
The above results show that the optimal scheduling scheme should strike a good balance between the averaged resource usage and the scheduling conflict. 
The model-wise scheduling generates the smooth resource usage pattern and hence low conflict, but uses unnecessarily more resources to meet the QoS target.
In contrast, layer-wise scheduling uses the minimum CPU resources, but the per-layer characteristics lead to a substantial scheduling conflict overhead.
The layer block scheduling combines the advantages of both model-wise and layer-wise scheduling.
Furthermore, the optimal performance cannot be achieved by simply setting a fixed layer block size as demonstrated in \Fig{fig-GrainPerf}. 
In other words, the optimal block organization depends on the model characteristics and query arrival rate.
As such, we propose to use adaptive layer block-size, and will explain it with greater details in later sections.

\subsection{Compilation Strategy Analysis}
\label{subsec:compile_analysis}

We now perform a set of experiments to study the impact of compilation strategies on multi-tenant deep learning services.
The key insight in our experiments is that the optimal compilation strategy changes under different interference levels.
As such, the adaptive compilation is needed to achieve high-performance multi-tenant DL services.
Furthermore, we propose to use multi-version static compilation to avoid the overhead of just-in-time (JIT) compilation.


\paragraphbold{Extending TVM Auto-Scheduler.}
Recall that in \Sec{subsec:compilation}, the current TVM compilation strategy uses an auto-scheduler~\cite{Ansor} to search for the implementation that achieves the best or the lowest latency.
This compilation strategy does not consider the existence of interference when multiple DNN models run together, which can lead to a significant performance slowdown as shown in \Fig{fig-InterferenceSnapshot}b.

To mitigate the impact of interference, we propose a naive extension for the TVM's existing auto-scheduler~\cite{Ansor}.
To identify the best code implementation for the target layer at a given interference level, we launch a background layer that produces the desired level of interference and run the TVM's auto-scheduler with long enough iterations (e.g., 1024 iterations).
As such, the returned schedule can be regarded as the optimal version under this interference level.
In this experiment, we use a frequently occurred ResNet \texttt{conv} layer with the feature map size of $14\times 14$, kernel size of $3\times 3$, input and output channel size of $256$, and study the performance of different compilation strategies under different interference levels.


\paragraphbold{Results.}
\Fig{fig-VerBehaviourAndLossRegion} compares the performance of four different implementations under different interference levels.
In specific, the four implementations correspond to the optimal ones searched with zero, low, medium, and high interference levels, respectively.
As \Fig{fig-VerBehaviourAndLossRegion}a shows, the \texttt{impl.-1}, which is also the default choice of TVM auto-scheduler, achieves the best performance when no interference exists.
However, its performance also degrades rapidly, which can be up to $7\times$ at the high inference level.
In contrast, the \texttt{impl.-4} has the lowest performance when no inference exists but achieves the highest performance under the high interference.
These results show that the optimal code implementations vary according to the interference levels, and our simple extension to the TVM auto-scheduler can effectively find these optimal implementations.

Since a model may experience all ranges of interferences in the multi-tenant DL services at one run, a static code version cannot achieve the best performance.
\Fig{fig-VerBehaviourAndLossRegion}b further quantifies the performance trend of the above four versions against different interference levels, where each version outperforms others only within a narrow interference interval.
As such, we have to combine all the four versions across all the interference levels to achieve the best performance, which is the dotted grey line in \Fig{fig-VerBehaviourAndLossRegion}.

\paragraphbold{General Cases.}
We further profile the rest of the ResNet-50 layers under different interference levels to fully understand the impact of the compilation strategy.
Specifically, we choose ten interference levels and identify the best-performing version at each level, which leads to a total number of ten implementation versions for each layer.
\Fig{fig-VerLossAndVerNum}a compares the performance loss of using a various number of versions against using all the ten versions.
If we use only one implementation, the performance loss increases as the interference level increases.
In contrast, using five versions out of the ten versions can maintain the performance loss within 10\%.

\paragraphbold{Multi-Version Static Compilation.}
One naive way to exploit the above insights for multi-tenant DL services is to perform a Just-in-Time (JIT) compilation according to the interference level.
However, the JIT compilation overhead can offset the benefit of adaptive compilation.
Instead, we propose to use the static multi-version compilation to achieve the same benefit of the adaptive JIT compilation.
\Fig{fig-VerLossAndVerNum}b plots the ratio of code version count to maintain various performance losses compared to the case of using all the ten versions.
Although the above results have shown that it requires five code versions to stay within 10\% performance loss, the majority (i.e., over 80\%) of layers only require three code versions.

%% file: DesignAndImplementation.tex
\begin{figure*}
	\centering
    \includegraphics[width=0.245\linewidth]{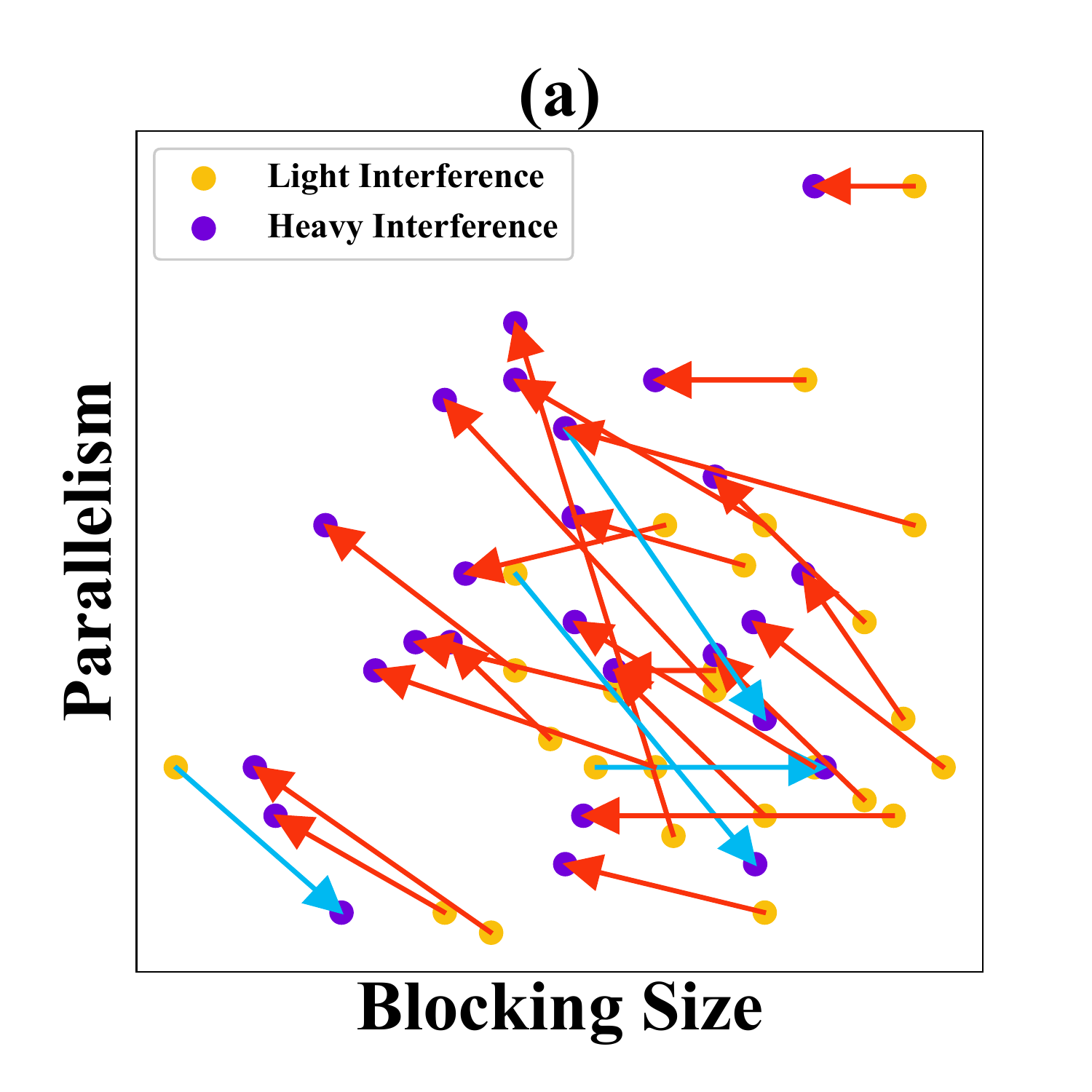}
    \includegraphics[width=0.245\linewidth]{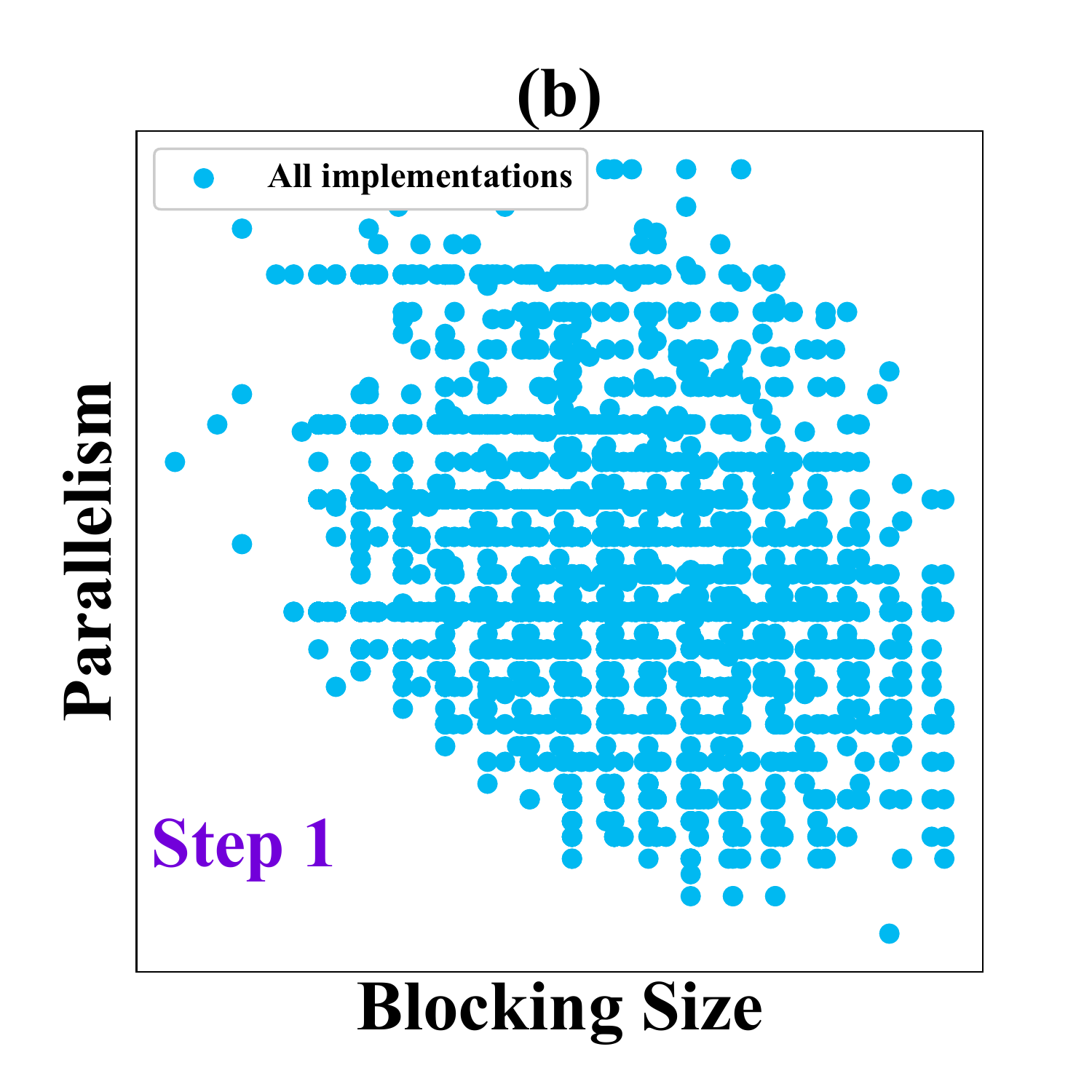}
    \includegraphics[width=0.245\linewidth]{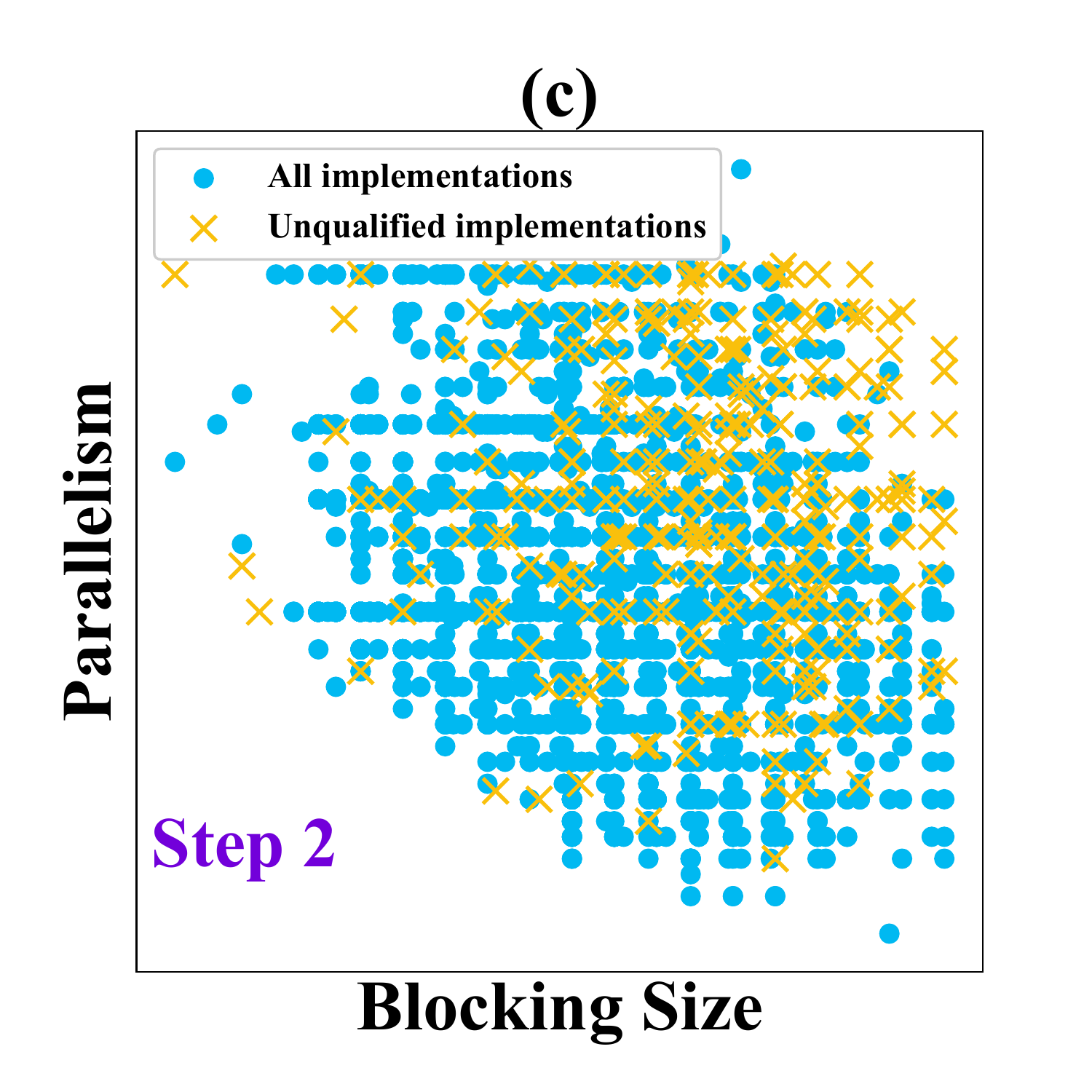}
    \includegraphics[width=0.245\linewidth]{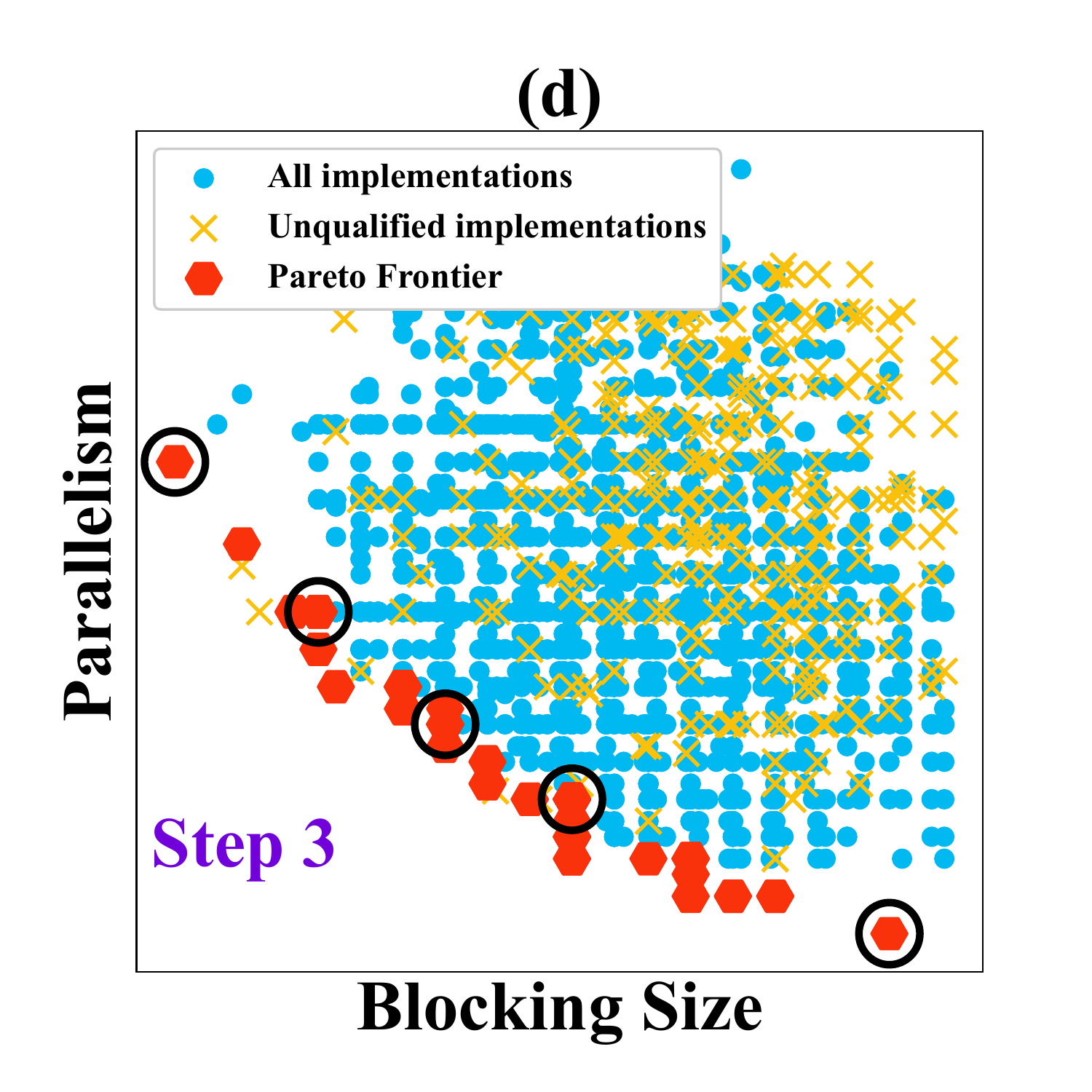}
	\caption{
	(a) The heavy-inference-optimal version generally prefers a large parallelism and a small blocking size (low locality), while the light-inference-optimal version prefers the opposite.
	(b-d) Steps to find optimal code version under different interference levels. We use an exemplary \texttt{conv} layer of $H_{in}=W_{in}=7$, $C_{in}=832$, $C_{out}=384$, $H_{K}=W_{K}=1$, $H_{out}=W_{out}=7$.}
	\label{fig-VersionShifting}
\end{figure*}

\section{Detailed Design of \proj{}}
\label{sec:Methodology}

In this work, we propose \proj{}, a software solution for high-performance multi-tenant deep learning model serving.
Based on the previous insights, \proj{} performs adaptive compiling and scheduling.
\Fig{fig-workflow} shows its overview that has two main components, i.e., the offline static compiler and the online runtime scheduler.

Instead of performing dynamic compilation whose overhead can account for the model serving latency, we propose to use a static multi-version compiler that extends the existing TVM's compilation framework and can identify different optimal code versioned under different interference levels. 
The key novelty in our compiler is a single-pass multi-version search algorithm (described in \Sec{subsec:adaptive_compilation}).

The \proj{} runtime scheduler dynamically forms the layer block as the scheduling unit, which balances the core usage efficiency and scheduling conflict rate.
The key in the scheduler is a dynamic threshold-based layer block formation algorithm that we describe in \Sec{subsec:adaptive_scheduling}.
The runtime scheduler exploits a performance counter-based interference proxy.
By monitoring the system load and interference pressures, it adaptively selects the optimal code version and scheduling granularities, which are detailed in \Sec{subsec:scheduler}.



\input{FindingImplementations.tex}

\input{FormingSubmodel.tex}

\input{Workflow.tex}

%% file: FindingImplementations.tex
\subsection{Single-pass Static Multi-version Compiler}
\label{subsec:adaptive_compilation}

In \Sec{subsec:compile_analysis}, we have described the benefits of adaptive compilation for handling the interference in the multi-tenant DL services.
We have proposed a naive extension for the TVM auto-scheduler to search for the best code version at a given interference level.
The extended auto-scheduler launches an additional background layer that can generate the desired interference level during the search process.
This approach is effective for identifying the best code version at different interference levels but is time-consuming as it requires multiple passes of the TVM auto-scheduler.
A single pass for a layer is typically 20 minutes on our high-end CPU, which means searching for five versions would take close to two hours.

To facilitate the multi-version compilation process, we propose a single-pass search algorithm that adds almost no overhead to the original TVM auto-scheduler.
Our key insight is that we can use the well-known computer architecture tradeoff between parallelism and locality to explain why certain versions are extremely sensitive to interference while others are much less sensitive.
Built upon this insight, we can explore the parallelism-locality tradeoff space in a single search pass, from which we then pick the desired versions.



\paragraphbold{Parallelism-Locality Tradeoff.}
We first use experimental results to illustrate that the finding of different optimal versions under different interference levels is essentially a tradeoff between program parallelism and locality.
In this experiment, we use the straightforward extension described in \Sec{subsec:compile_analysis} to search for the two optimal code versions, one under the light interference level and the other one under the heavy interference level.
We then record the corresponding complication flags for these two versions.
Based on the recorded flags, we compute a parallelism metric by simply multiplying the loop unrolling factor and parallelization factor.
We compute a locality metric by directly using the tiling/blocking size.

In \Fig{fig-VersionShifting}a, we compare the above two metrics of the two code versions that achieve the best performance under the light inference and heavy inference, respectively.
We observe that the heavy-inference-optimal version generally prefers high parallelism and a small blocking size, while the light-inference-optimal version prefers the opposite.
We then derive the following insight:
generated codes with a higher locality (a larger blocking size) perform better under the light interference (\textit{\textbf{interference-vulnerable}}), while generated codes with a higher parallelism perform better under the heavy interference (\textit{\textbf{interference-tolerant}}).
 
\begin{algorithm}[b]
	\caption{\textbf{Static multi-version compilation in a single pass.}}
	\label{Algo-FindingVersion}
	\begin{flushleft}
	\textbf{Input:}  $ layers\lbrack N \rbrack $, $ qos $\\
	\textbf{Output:}  $ candidate\_impls\lbrack N \rbrack\lbrack V \rbrack $, $ dominant\_impls\lbrack N \rbrack\lbrack  \rbrack $
	\end{flushleft}
	\begin{algorithmic}[1]
		\Function {FindingImpl}{$ layers $, $ qos $}
        \For {$ \_l $ \textbf{in} $ layers \lbrack N \rbrack$}
        \State {$ \_l.qos\leftarrow qos\times \frac{\_l.op\_count}{\sum_{x\in layers\lbrack 1:N \rbrack}(x.op\_count)} $}
        \State {$ impls\lbrack  \rbrack\leftarrow Ansor(\_l, 1024) $}
        \State {$ impls\lbrack  \rbrack\leftarrow \lbrack x.time\leq \_l.qos~for~x~in~impls \rbrack $}
        \State {$ d\_impl\lbrack  \rbrack\leftarrow ExtractDominant(impls) $}
        \State {$ d\_impl\lbrack  \rbrack.sort(key=x.block\_size) $}
        \For {$ \_i $ \textbf{in} $ d\_impl, step\leftarrow\frac{d\_impl.length}{V} $}
        \State {$ c\_impl.push\_back(\_i) $}
		\EndFor
		\State {$ candidate\_impls.push\_back(c\_impl) $}
		\State {$ dominant\_impls.push\_back(d\_impl) $}
        \EndFor
        \State \Return {$ candidate\_impls $, $ dominant\_impls $}
		\EndFunction
		
	\end{algorithmic}
\end{algorithm}


The above insight reflects the well-understood parallelism-locality tradeoff.
To exploit the large locality, a layer needs to use more on-chip memory like the LLC in CPU, which are shared resources among multiple CPU cores.
However, the performance of the layer quickly degrades when there is contention on the shared resources (L3 cache and the corresponding bandwidth according to our observation).
To mitigate the impact of contention, the layer can limit its locality and use more parallelism to remedy its performance loss.


\paragraphbold{Single-Pass Compilation.}
We now use the examples in \Fig{fig-VersionShifting}b-d to walk through our single-pass compilation algorithm, which has three steps. The details of the algorithm are provided in \Agl{Algo-FindingVersion}.

The first step (Line 2 - 4 in \Agl{Algo-FindingVersion}) directly leverages the TVM's auto-scheduler to collect candidate implementations.
In this step, we enable the operator fusion optimization in the auto-scheduler, which includes  common fusion patterns like convolution followed by ReLU (\texttt{conv-relu}) and convolution followed by batch normalization and ReLU (\texttt{conv-batchnorm-relu}).
Instead of searching for the best-performing implementation, we record as many samples as possible and calculate their parallelism and locality metrics as \Fig{fig-VersionShifting}b shows.
In the second step (Line 5), we then filter out samples whose performance can not satisfy this layer's QoS target as \Fig{fig-VersionShifting}c shows.
We set the layer's performance as the minimal floating-point operation per second (i.e., FLOPS) that the corresponding model needs to achieve to meet the model's latency target.

In the third step (Line 6 to 7 and Line 14 to 29), we select the \emph{dominant} implementations via $ ExtractDominant$ function, where there are no other implementations with both smaller blocking size and parallelism than each chosen one. 
In other words, these dominant implementations form the Pareto frontier (red markers in \Fig{fig-VersionShifting}d), which is an optimal solution to the multi-objective optimization problem.
In the last step (Line 8 to 12), we uniformly choose five versions from the Pareto frontier (circled ones in \Fig{fig-VersionShifting}d).
Since not all layers require five versions to maintain the close performance to the optimal, we test the performance of the selected five versions under different interference levels and remove the ones whose performance is within 90\% of the full five versions.
This optimization leads to the reduced storage overhead of code multi-versioning.



%% file: FormingSubmodel.tex
\subsection{Dynamic Threshold Based Layer-Block Formation}
\label{subsec:adaptive_scheduling}

As previously explained in \Sec{subsec:scheduling_analysis}, the layer-block-based scheduling outperforms the layer-wise and model-wise scheduling through balancing the minimal average core usage and scheduling conflict rate.
However, a fixed-sized layer-block is not efficient because the optimal block size varies with the system load and the interference from other co-executed models.
As such, we propose a \emph{dynamic-sized} layer-block approach to achieve the high core efficiency and low conflict rate according to the system load and interference level.



\begin{figure}[t]
    \centering
    \includegraphics[width=0.495\linewidth]{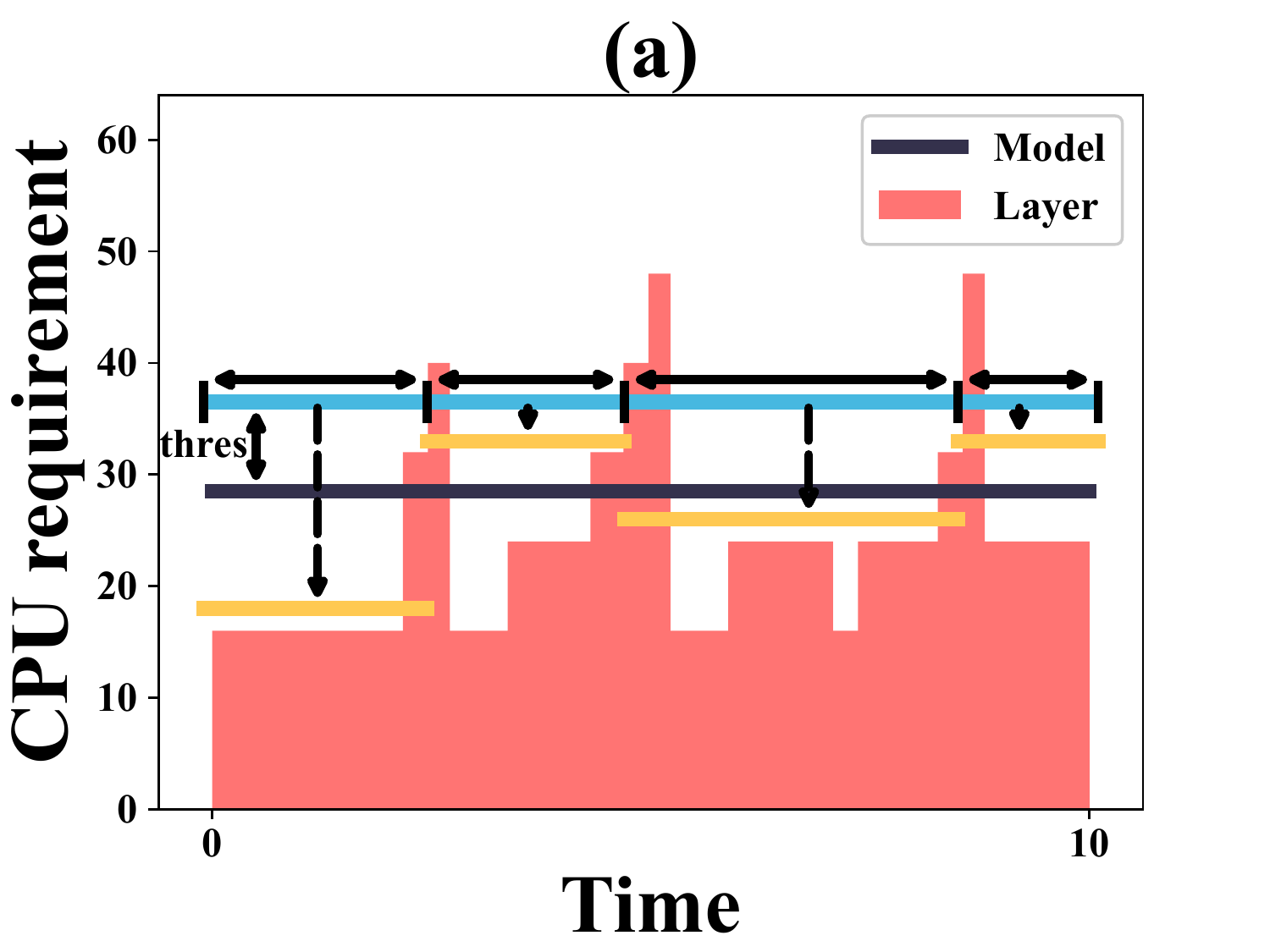}
    \includegraphics[width=0.495\linewidth]{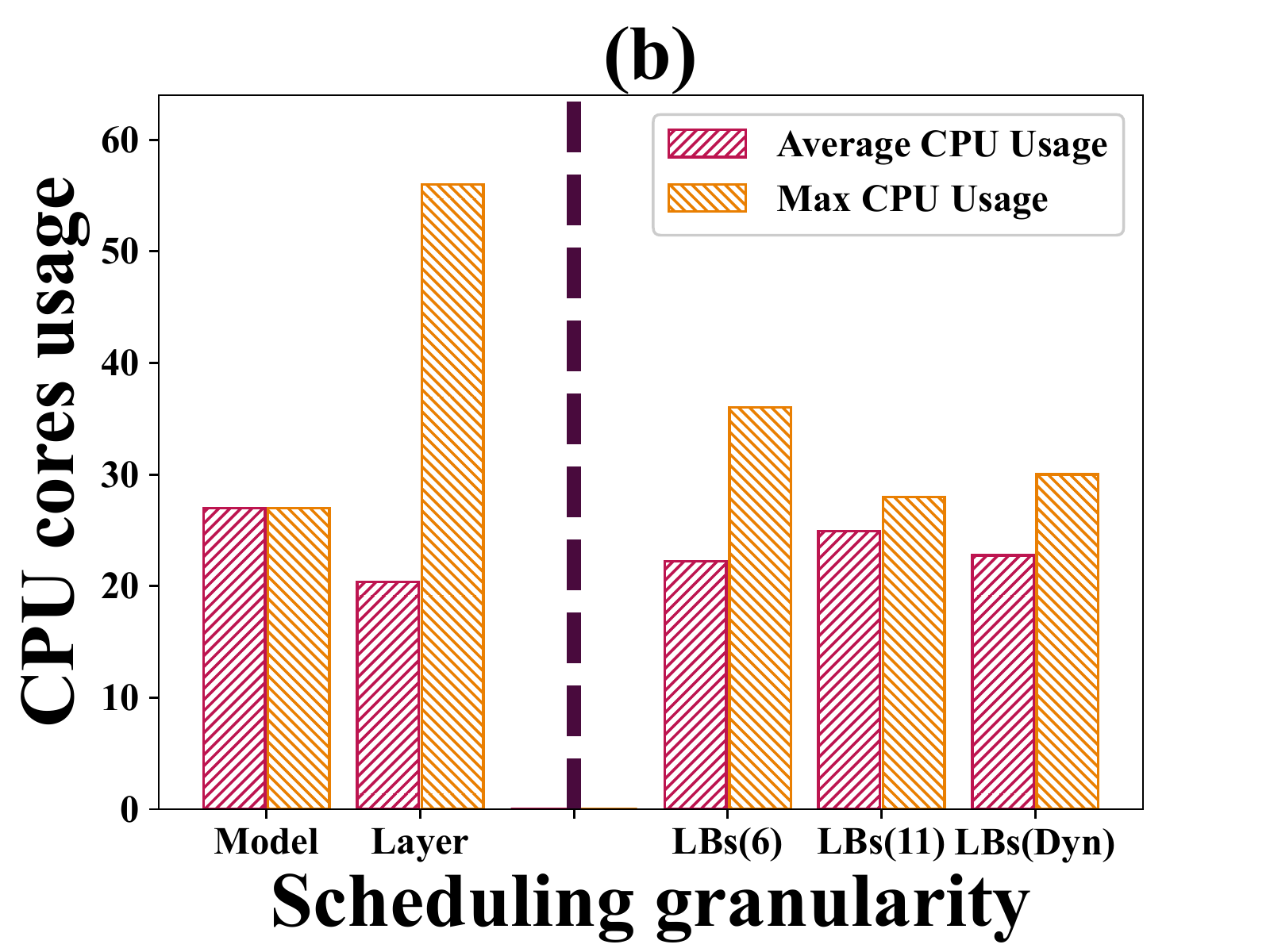}
    \caption{\label{fig-FormingVeltairSM} (a) Forming the layer-blocks by a threshold and minimize the layer-blocks' CPU usage. (b) Average and maximal CPU usage of various scheduling granularity.}
\end{figure}

To reduce the scheduling conflict rate via the layer-block scheduling, we first identify the layers that are most likely to trigger conflicts.
To identify these conflict-prone layers, we calculate the required CPU core number for each layer to complete within its QoS target, from which we can compute the model's averaged core number.
We compare the layer-wise core number against the model-wise average value, and identify the layers with a much higher CPU core number requirement than the averaged value as conflict-prone layers.
For each conflict-prone layer, we form a layer-block that can reduce its core usage by increasing the core usage for other layers in the block while still satisfying the QoS target.

We walk through the ResNet example in \Fig{fig-FormingVeltairSM}a to illustrate the intuition of our method.
We first form the layer-wise (red shadowed area) and model-wise (black horizontal line) scheduling plan in \Fig{fig-FormingVeltairSM}a.
We denote the average core count in the mode-wise scheduling as $ Avg\_C $.
We then use a runtime-decided threshold $ thres $ that ranges from zero to maximal core count.
We iterate over all layers and identify the conflict-prone ones whose core requirement exceeds $ Avg\_C+thres $, i.e., the blue line in \Fig{fig-FormingVeltairSM}a.
We refer to each conflict-prone layer to the splitting pivot, which is essentially the beginning layer for a block.
As a result, there are four blocks marked by arrows.
For each formed block, we calculate its QoS target by summing up all its layers.
We then recalculate the core requirement (yellow line) of each block to satisfy its QoS target.




\begin{algorithm}[b]
    \caption{\textbf{Dynamic threshold based layer-block formation algorithm.}}
    \label{Algo-FormingSubModel}
    \begin{flushleft}
    \textbf{Input:}  $ layers\lbrack N \rbrack $, $ impls $, $ thres $\\
    \textbf{Output:}  $ Layer\_Block $
    \end{flushleft}
    \begin{algorithmic}[1]
        \Function {Finding1stPivot}{$ layers $, $ impls $, $ thres $}
        \State {$ splitting\_pivot\leftarrow 0 $}
        \State {$ Avg\_C\leftarrow Core_{@~Model~Granularity}(layers) $}
        \For {$ \_l $ \textbf{in} $ layers[1:N]$}
        \If {$ Core(impls[\_l])\geq thres+Avg\_C $}
        \State {$ splitting\_pivot\leftarrow \_l $}
        \State {\textbf{break}}
        \EndIf
        \EndFor
        \State \Return {$ splitting\_pivot $}
        \EndFunction
        \Function {FormingLBs}{$ layers $}
        \State {$ Layer\_Block\leftarrow[], begin\leftarrow 0 $}
        \While {$ layers.length() \neq 0$}
        \State {$ sp\leftarrow Finding1stPivot(layers, impls, thres) $}
        \State {$ Layer\_Block.push\_back(layers[begin:sp]) $}
        \State {$ layers\leftarrow layers[sp+1:] $}
        \State {$ begin\leftarrow sp+1 $}
        \EndWhile
        \State \Return {$ Layer\_Blocks $}
        \EndFunction
    \end{algorithmic}
\end{algorithm}

\Agl{Algo-FormingSubModel} formally describes the above dynamic threshold-based layer-block formation algorithm.
 where the threshold is determined at runtime according to the system load and co-executed models' characteristics.
\Sec{subsec:scheduler} will provide the details of how we adjust the threshold.
With \Agl{Algo-FormingSubModel}, we can generate proper layer-blocks using no more than $ Avg\_C+thres $ CPU cores under different system loads.
The basic idea is that when the system load is low, we use a high threshold since the conflict possibility is low, which means each layer can use as many cores as possible for maximizing the CPU resource usage efficiency.
When the system load is high, we use a low threshold which means each layer should have core counts close to the average value for reducing the scheduling conflict rate. 

The essence of \Agl{Algo-FormingSubModel} is to reduce the high core usage of error-prone layers and remedy its latency loss by increasing the core usage of other layers.
This can increase average core usage compared to the most efficient layer-wise scheduling.
However, according to our evaluation, the gap between optimal CPU resource usage is smaller than $ 10\% $, which is acceptable.
\Fig{fig-FormingVeltairSM}b compares the average core usage and maximum core usage of different scheduling granularities when co-locating two ResNet-50 models.
Our algorithm is effective at reducing the gap between optimal CPU cores usage (i.e., achieving high resource efficiency) and maximum core usage (i.e., reducing the scheduling conflict rate).

%% file: Workflow.tex
\begin{figure}[t]
    \centering
    \includegraphics[width=0.495\linewidth]{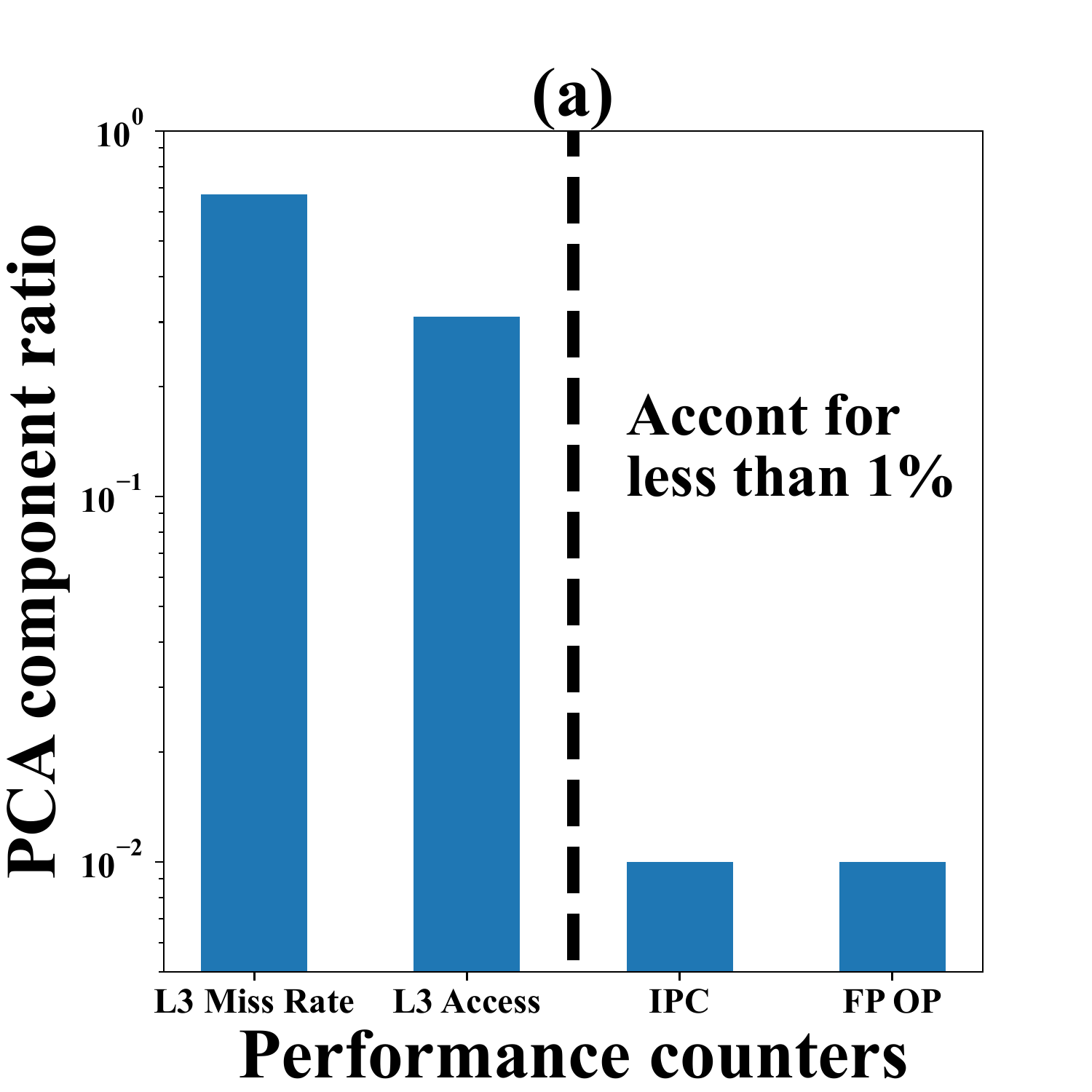}
    \includegraphics[width=0.495\linewidth]{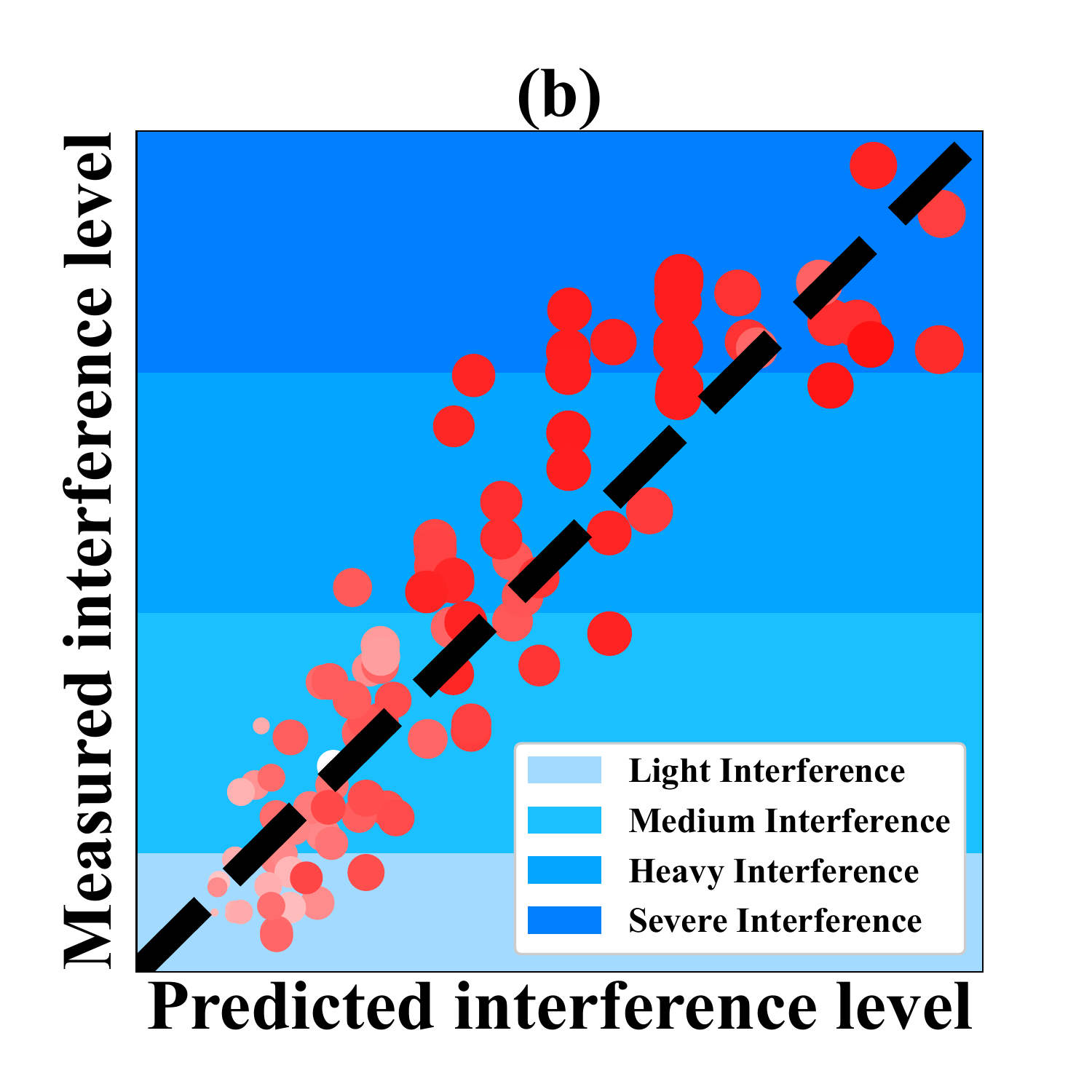}
    \caption{\label{fig-Verify} (a) Dominant components of performance counters according to PCA. (b) Accuracy validation of the interference pressure proxy using L3 miss rate and access counters.}
\end{figure}

\subsection{\proj{} Runtime Scheduler}
\label{subsec:scheduler}

In this subsection, we describe the details of the runtime scheduler in \proj{}.
Our scheduler monitors the current CPU inference level and dynamically chooses the code version derived from \Sec{subsec:adaptive_compilation} and scheduling granularity using the algorithm described in \Sec{subsec:adaptive_scheduling}. 
In specific, we explain how we derive the system interference pressure level and how we determine the dynamic threshold for \Agl{Algo-FormingSubModel}.

\paragraphbold{Interference Proxy.}
We first build the proxy for monitoring the system interference pressure level by using hardware performance counters.
According to previous studies~\cite{Protean,Themis,HSM}, performance counters have a strong relation with interference pressure level. We define the interference pressure level of the system as the average performance slowdown ratio of layers running on the system.
To figure out what performance counters decide the interference level, we conduct a principal component analysis (PCA)~\cite{PCA} on collected performance counters, including L3 cache miss Rate, L3 access, instruction per cycle (IPC), float-point operations, etc. 
It turns out that L3 cache-related counters account for over 99\% of the data variance, as shown in \Fig{fig-Verify}a. 
As such, we choose the L3 miss rate and L3 access to construct a simple linear interference model.
As shown in \Fig{fig-Verify}b, the predicted interference level matches the measured interference level well. 
Using this simple linear model, we can derive the interference pressure level with low cost at runtime.


\begin{algorithm}[b]
    \caption{\textbf{The details of \proj{} scheduler.}}
    \label{Algo-VeltairScheduler}
    \begin{algorithmic}[1]
    \Function{VeltairTaskDispatcher}{}
    \State {\textbf{Dispatch tasks following $ Poisson $ distribution}}
    \EndFunction
    \Function{VeltairWorker}{}
    \While {true}
    \If {worker is busy}
    \State {Wait for last task to finish}
    \EndIf
    \State {$ t\leftarrow fetch\_task(), begin\leftarrow 0 $}
    \While {$ t.finished\neq True $}
    \State {$ i\leftarrow system~interference $}
    \State {$ thres\leftarrow \#C_{Total}-$$\sum_{t_{active}}(\#C_{Model~Granularity}(t)) $}
    \State {$ pivot\leftarrow Finding1stPivot(t, impls_{i}, thres) $}
    \State {$ t[begin:pivot].Execute() $}
    \State {$ t\leftarrow t[pivot+1:] $}
    \State {$ begin\leftarrow pivot+1 $}
    \EndWhile
    \EndWhile
    \EndFunction
    \end{algorithmic}
\end{algorithm}

\paragraphbold{Dynamic Scheduling Threshold.}
The threshold used by the layer-block formation in \Sec{subsec:adaptive_scheduling} indicates the additional core counts that each layer block can use beyond the model's averaged requirement.
When a model runs exclusively, it can use as many cores as it desires.
However, when multiple models run concurrently, each model should try to reduce its core usage to avoid scheduling conflicts.
As such, we use a simple heuristic that determines the threshold by subtracting the total core number by the sum of all models' average core count and distributing the remaining cores according to each model's average core count.
For example, three models \texttt{A}, \texttt{B}, \texttt{C} use $ 12, 12, 24 $ CPU cores on average respectively.
On average, $ 64-(12+12+24)=16 $ cores are idle, and we assign $ 4, 4, 8 $ as threshold to \texttt{A}, \texttt{B}, \texttt{C} respectively.
In our study, we observe that a model with a high average core usage typically has a high peak core usage.
Thus, dividing the idle cores by the model's average core usage can better fit each model's computation demand.


\begin{figure*}[t]
        \centering
    \includegraphics[trim=0 .7cm 0 0, clip,width=\linewidth]{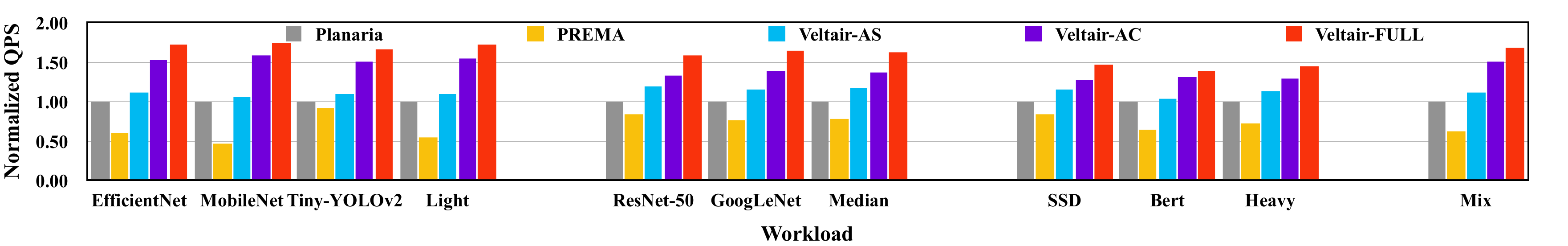}
        \caption{\label{fig-evalA} 
        Query per second (QPS) with 95\% tasks QoS satisfied for various workloads and scheduling strategies.}
        \vspace*{0.2cm}
\end{figure*}

\paragraphbold{Putting All Together.}
We now describe the runtime scheduler in \proj{} that exploits the aforementioned algorithms. 
\Agl{Algo-VeltairScheduler} illustrates the pseudo-code of the scheduler, where the task dispatcher simply sends tasks to a worker if there are enough idle cores. 
The code implementation search is done offline at the compiling stage as  \Agl{Algo-FindingVersion} details. 
At runtime, \proj{} collects the performance counters once a layer block is finished, and the scheduler will form the next layer block from the remaining layers according to the current system load with different implementations according to the current interference pressure level. 
As mentioned before, codef implementations with different interference tolerance levels have significant differences in parallelism and locality, which means the same layer in a model will have different CPU requirements under different interference pressure levels leading to different layer blocks. 
Note that when selecting the interference tolerance level of the next layer-block, we will ignore the ongoing but soon-to-finish layer-blocks, since they will have little influence on system interference from now on. 
To determine the soon-to-finish layer-block, we examine whether its remaining execution latency is within a threshold (e.g., 10\%) according to our offline profile-based latency model and interference proxy model.

%% file: Evaluation.tex
{
\begin{table}[b]
\setlength{\arrayrulewidth}{1.1pt}
\renewcommand{\arraystretch}{1.3}
        \caption{Evaluated multi-tenant DL models.}
        \label{tab-Workload}
        \centering
        \resizebox{\linewidth}{!}{
        \begin{tabular}{|c|c|c|c|}
                \hline
        Category & Workload & Name & QoS (ms)\\
        \hline
                        & Medium & ResNet-50~\cite{ResNet} & 15 \\
        Image           & Medium & GoogLeNet~\cite{GoogLeNet} & 15 \\
        Classification  & Light & EfficientNet~\cite{EfficientNet} & 10 \\
                        & Light & MobileNet-V2~\cite{MobileNetV2} & 10 \\
        \hline
        Object          & Heavy & SSD~\cite{SSD} & 100 \\
        Detection       & Light & Tiny-YOLOV2~\cite{Tiny-YOLOV2} & 10 \\
        \hline
        NMT             & Heavy & Bert-Large~\cite{Bert} & 130 \\

                \hline
        \end{tabular}
        }
\end{table}
}

\section{Evaluation}
\label{sec:Evaluation}
We now demonstrate the effectiveness of the adaptive compiling and scheduling in \proj{}. We first describe our evaluation setup, baseline, and metrics.
We compare the performance of \proj{} against other work and show that the combination of our adaptive compiling and scheduling is essential for achieving the hight-performance in multi-tenant DL services.

\subsection{Experimental Setup}

\paragraphbold{Multi-Tenant Deep Learning Models.}
To simulate the realistic situation of deep learning services, we use deep learning models from MLPerf (Server)~\cite{MLPerf} as listed in \Tbl{tab-Workload}.
The evaluated models include image classification, object detection, and neural machine translation (NMT) tasks.
We categorize the workload of the models from light, medium to heavy, and set the QoS target for them according to the guidance of MLPerf.
 
\paragraphbold{Workload Generation.}
We also follow the MLPerf guidance to generate random queries with Poisson distribution, where the $ \lambda $ parameter of the distribution stands for the QPS (query per second) of the workload. We evaluate our design under \textit{Light}, \textit{Medium}, \textit{Heavy}, and \textit{Mix} workload.
For the mixed workload, the frequency of every task is set to be inversely proportional to QoS requirements~\cite{GoogleWorkload}. 

\paragraphbold{Hardware and Software.}
For all experiments, we use a machine equipped with a high-end server-level CPU Ryzen Threadripper 3990X~\cite{3990X} and 256~GB DDR4 RAM at 3200~MHz.
The CPU has 64 physical cores and 256MB L3 cache capacity, and works at 2.9~GHz with AVX-2 enabled.
To obtain stable experimental results, we turn off certain features such as simultaneous multi-threading (SMT) and dynamic voltage and frequency scaling (DVFS).
We believe that turning off these features do not change our insights owing to the following reasons. 
The SMT mainly enhances the sharing of L1 cache, while we identify LLC as the main contentious resource. 
However, SMT leads to a significant latency fluctuation because of the possibility that two logical threads of different tasks are assigned to the same physical core. 
The DVFS also leads to latency fluctuation that increases the conflict rate.
We implement the static multi-version compiler by extending the TVM v0.8~\cite{TVM}.
For the runtime scheduler implementation, we use MPICH 3.3.2~\cite{MPI}, which serves the multi-tenant DNN models via multi-processing.

\paragraphbold{Evaluation Metrics.}
We use QPS with 95\% tasks QoS satisfied, average latency, and CPU usage efficiency as our evaluation metrics.
\begin{itemize}
    \item \underline{QPS with 95\% Tasks QoS Satisfied}: this metric represents how many requests the system can serve per second with almost all the query requests (95\%) finish within the QoS target.
    \item \underline{Average Latency}: This metric measures the average execution latency of all the queries. 
    \item \underline{CPU Usage Efficiency}: This metric measures the average CPU usage of the tasks by dividing the total execution time by the sum of multiplying of the core usage and execution time of each layer.
\end{itemize}

\paragraphbold{Baseline Choice.}
Since we co-locate multiple DNN models and let them spatially share the hardware, we choose Planaria~\cite{Planaria} as the baseline in our evaluation.
It should be noted that Planaria is based on the hardware-software co-design, while we port the software scheduling part to the CPU platform.
To justify why we only consider the spatial multitasking scenario, we also implement another baseline scheduling method PREMA~\cite{PREMA}, which is a temporal multitasking algorithm and lets tasks with high priority preempt.  

\paragraphbold{Evaluation Plan.}
We study the effectiveness of different components by evaluating the following configurations of \proj{}.
\begin{itemize}
 \item \underline{\proj{}-AS}: with only adaptive scheduling.
 \item \underline{\proj{}-AC}: with only adaptive compilation.
        \item \underline{\proj{}-FULL}: with both adaptive scheduling and adaptive compilation enabled.
\end{itemize}

\subsection{Query per Second (QPS) Improvement}
\Fig{fig-evalA} demonstrates the QPS improvement of \proj{} against the baseline Planaria~\cite{Planaria} for studied models in different levels of workloads. 
\proj{}-FULL achieves an average of 71\%, 62\%, 44\% improvement in the light, medium, heavy workloads respectively, and an average of 68\% improvement in the mix workloads. 

We also observe that the adaptive compilation (\proj{}-AC) achieves better improvements than the adaptive scheduling (\proj{}-AS).
However, these two techniques are synergistic, and both are critical components for fulfilling the performance improvement of the full version of our design (\proj{}-FULL).
Without the adaptive scheduling, \proj{}-AC only achieves 50\% QPS improvement in contrast to the 68\% improvement of \proj{}-FULL in the mix workloads. 
The reason is that without adaptive scheduling, many layers will choose implementations with lower locality but higher parallelism to handle interference, leading to increased CPU requirement and thus increased conflict possibility. 
In contrast, the dynamic layer block formation in adaptive scheduling can mitigate these conflicts.

In \Fig{fig-evalA}, we also observe that the temporal multitasking-based multi-DNN serving scheme (PREMA~\cite{PREMA}) generally performs worse than the spatial multitasking-based multi-DNN serving.
This observation justifies the choice of spatial multitasking of our work.

\subsection{Query Execution Latency Result}

\begin{figure}[t]
        \centering
    \includegraphics[trim=0 1.2cm 0 0, clip,width=\linewidth]{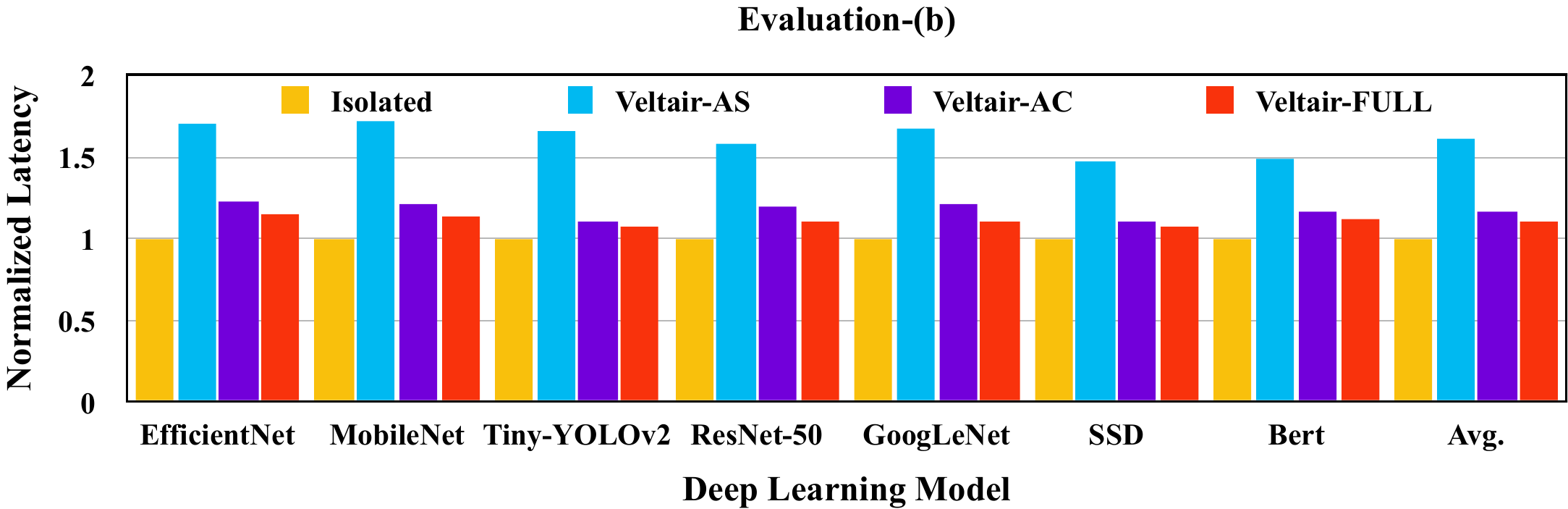}
        \caption{\label{fig-evalB} Average query execution latency comparison between solo-run (isolated) and various \proj{} configurations (\proj{}-AS, \proj{}-AC, \proj{}-FULL).}
\end{figure}

We compare the average query latency of various \proj{} configurations against the solo-run case in \Fig{fig-evalB}.
For each model, the latency is measured at the QPS where 95\% of queries can meet their QoS target (i.e., same to the QPS metric in \Fig{fig-evalA}).
Since the solo-run latency is the shortest latency each model can achieve on the studied CPU platform, this comparison lets us identify how much additional room there is for further optimization in \proj{}.

\Fig{fig-evalB} shows that the inference latency of \proj{}-AS is $ 1.6\times $ of the isolated solo-run execution, which means the adaptive scheduling cannot reduce the execution latency.
On the other hand, the latency of \proj{}-AC is $ 1.17\times $ of the isolated execution, confirming its ability for reducing latency under interference.
With both adaptive scheduling and compilation, the average latency is only $ 1.1\times $ of the isolated execution, which means that \proj{}-FULL is close enough to the optimal serving result on the studied platform.
 

\subsection{Result of CPU Efficiency}

In \proj{}, the layer-block-based scheduling leads to smoother CPU core usage with reduced conflict rate but potentially uses more cores. 
We quantify the gap of average core usage between \proj{} scheduling and the layer-wise scheduling.
Recall that the fine-grained layer-wise scheduling indicates the minimal core usage.
\Fig{fig-Sensitivity} shows that even under 75\% system load, the core usage gap of \proj{} is less than 10\% compared to the minimal core usage of the layer-wise scheduling.
In contrast, the model-wise has a much larger gap of 47\%.
These results confirm that our layer-block-based scheduling strikes a balance between reducing the scheduling conflict rate and maintaining the high resource usage. 


\subsection{Sensitivity and Overhead Analysis}

\begin{figure}[t]
        \centering
    \includegraphics[width=0.35\linewidth]{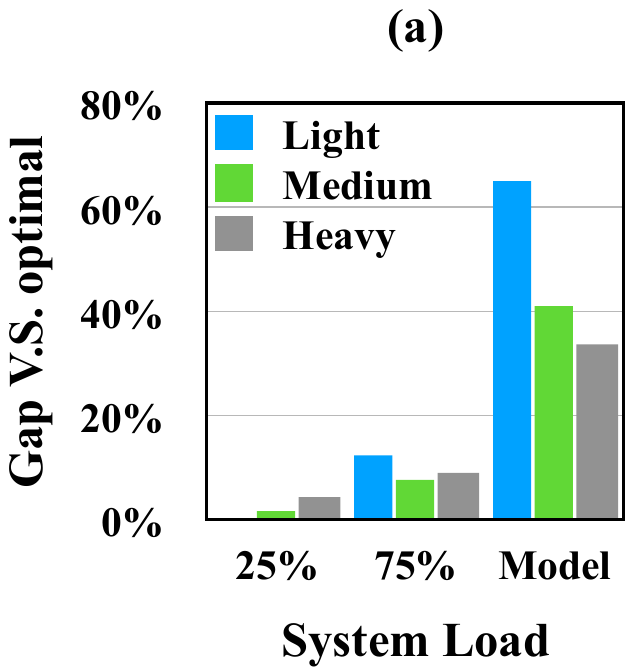}
    \includegraphics[width=3cm]{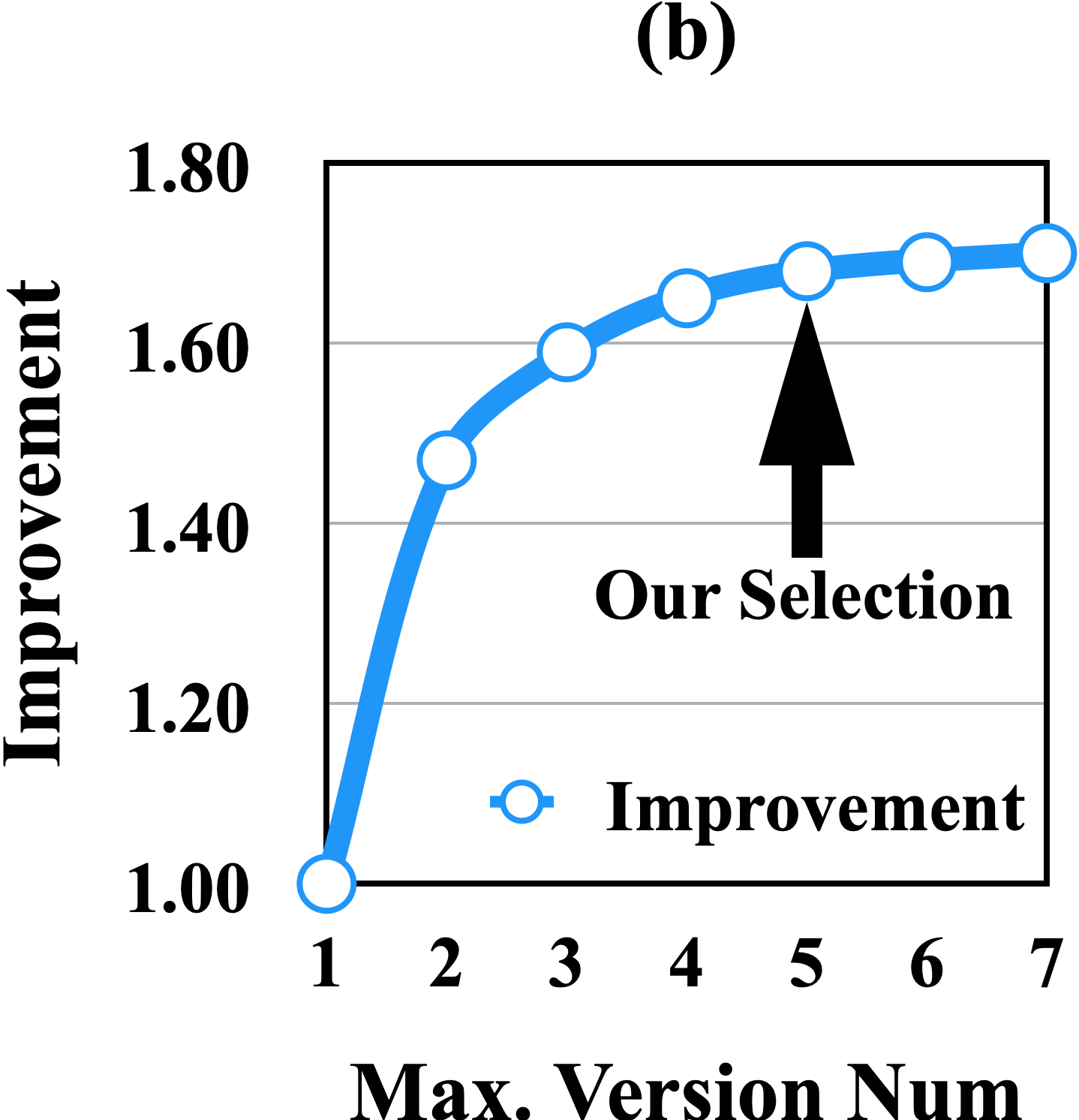}
    \includegraphics[width=0.26\linewidth]{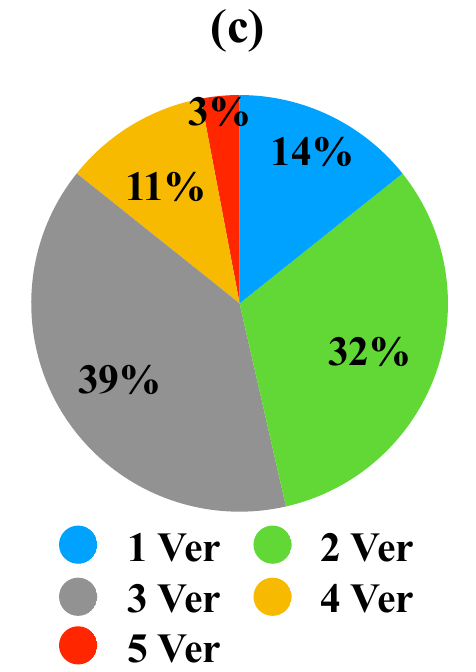}
        \caption{\label{fig-Sensitivity} (a) Gap between optimal core usage of layer block-based scheduling strategy under different system load, comparing to the fine-grained layer-wise scheduling. (b) Improvement under different version number. (c) The ratio of number of used versions for all the layers in seven DNN models.}
\end{figure}

\paragraphbold{Sensitivity.}
In \proj{}, we empirically set the maximal version number $ V $ to 5. 
We now study the performance improvement under different $ V $ in \Fig{fig-Sensitivity}b, which shows the improvement saturates after four versions.
\Fig{fig-Sensitivity}c plots the version count distribution for different layers, which shows that only 3\% layers require five versions.
These results justify the choice of using five versions.

\paragraphbold{Scheduling Overhead.}
The scheduling overhead of \proj{} mainly consists of two parts.
The first part is the runtime layer block formation procedure, which scans the layers only once and has the complexity of $ O(N) $.
The second part comes from the linear-model-based interference proxy. 
Owing to the low complexity of the scheduling algorithm and the proxy model, we find that their overall overhead is less than $0.1~ms$ for serving each DNN model.

%% file: RelatedWork.tex
\section{Related Works}
\label{sec:Related}

We compare and contrast \proj{} with previous works in the following three aspects, including the task co-location, multi-tenant deep learning services, and deep learning compiler.

\subsection{Task Co-location in Datacenter}
The rapid improvement of hardware computation power makes it possible to share the hardware among multiple tasks for higher throughput. 
Many works have studied how to co-locate a latency-critical (LC) task with multiple best-effort tasks ~\cite{Baymax, Heracles,Prophet}. 
Parties~\cite{Parties} proposes a resource-partitioning technique to co-locate multiple LC services. 
For more intelligent resource partition and management, CuttleSys~\cite{CuttleSys} proposes a sampling-reconstruction-prediction-based strategy with reconfigurable architecture. 
Bubble-series~\cite{BubbleUp, BubbleFlux} proposed an online contention measurement and control system to relax the performance loss caused by contention and is free from the online compiler, execution checkpoint, code variants rerouting, etc.
While previous works mainly focus on resource partition, isolation, and management, Protean~\cite{Protean} and other works~\cite{CompilingForNiceness} extend the optimization space by introducing runtime code transformation for lower L3 cache pollution. 

\subsection{Multi-Tenant Deep Learning Service}
Different from conventional workload, deep learning services are computation-intensive with complex inner structures and should be specifically treated when co-locating them. 
DART~\cite{DART} proposes a pipeline-based method to co-locate multiple DNN workloads on multiple heterogeneous computation nodes. 
While in this work, we mainly consider co-locating multiple DNN tasks on one homogenous hardware. 
PREMA~\cite{PREMA} and AI-MT~\cite{AI-MT} propose temporal multiplexing architectures with preemption-based strategy and computation-memory overlapping-based strategy, respectively.
In contrast, Planaria~\cite{Planaria} proposes a spatially decomposable systolic architecture to co-locate tasks with proper computation and memory resources. 
This work aims at relaxing the problem from compiling and scheduling aspect with less constraint on the back-end hardware as long as it is programmable. 
In addition to hardware-software co-design, DyNet~\cite{DyNet} mainly handles the problem of scheduling RNNs. LazyBatch~\cite{LazyBatch} proposes a batch-based approach to handle multiple DNN requests. Ebird~\cite{cui2019EBird} also proposes a batch-based approach to enable concurrent execution of DNNs with high data transfer-compute overlapping. Abacus~\cite{cui2021Abacus} proposes an operator overlapping strategy based on precise latency prediction. 
Besides the multi-DNN serving scenario, emerging microservice-based workloads also have complex inner structures similar to DNN models~\cite{Deathstar}, to which our design may also be applied.

\subsection{Deep Learning Compiler}
For the better flexibility and performance of DNN model execution, recent researchers propose various DL compilers including TVM~\cite{TVM}, TensorComprehensions~\cite{FAIR-TC}, Tiramisu~\cite{Tiramisu}, TensorFlow-XLA~\cite{XLA}.
These DL compilers are often integrated with front-end optimizers like TASO~\cite{TASO} or Grappler~\cite{Grappler}.
Meanwhile, they also introduce domain-specific language to make it convenient for users to define their own computation. 
For the code generation optimization with a huge search space, researchers apply both machine-learning-based methods including AutoTVM~\cite{AutoTVM}, Ansor~\cite{Ansor}, FlexTensor~\cite{FlexTensor} and heuristic based methods including DLFusion~\cite{liu2020DLFusion} and Paleozoic~\cite{liu2020Paleozoic}.
These compilers target general-purpose hardware or DL accelerators, and generally outperform vendors-provided libraries. 
However, these works mainly focus on optimizing the performance of the stand-alone execution of DNN operators.
In contrast, we explore compilation optimization for co-locating multiple deep learning tasks, for which we show the interference-aware compilation is critical. 

%% file: Conclusion.tex
\section{Conclusion}
\label{sec:Conclusion}
In this work, we proposed \proj{}, a compiler-scheduler system for high performance multi-tenant deep learning service. By leveraging multi-version compiling and layer-block scheduling, we achieve $ 1.7\times $ system maximal QPS and reduce $ 50\% $ of the computation latency with little overhead. We first evaluate the proper scheduling granularity in deep learning tasks, and we propose a layer-block scheduling strategy with dynamically adjustable size to reduce the resource conflict. Then we study the compilation options and propose a single-pass multi-version compilation to handle the performance loss of interference caused by shared resource competition in multiple neural networks co-locating. We demonstrate the advantages of \proj{} in the aspects of improvement in QPS, QoS satisfaction rate, computation latency, and resource usage efficiency using the standard MLPerf Server test suite.

\section*{Acknowledgement}
This work was supported by the National Key R\&D Program of China under Grant 2021ZD0110104, the National Natural Science Foundation of China (NSFC) grant (U21B2017, 62072297, 61832006).
We thank the anonymous reviewers and our shepherd Prof. Xipeng Shen for their constructive feedback for improving the work. We also thank Zhanda Zhu, Zihan Liu, Yijia Diao, and Vega Jiang for the beneficial discussion and continuous support.